\documentclass[letterpaper,11pt]{article}
\usepackage[DIV12]{typearea}
\usepackage{longtable}
\usepackage{booktabs,colortbl}
\usepackage{multirow}
\usepackage{amsmath}
\usepackage{amssymb}
\usepackage{bbm}
\usepackage[utf8]{inputenc}
\usepackage{pdflscape}
\usepackage{xspace}
\usepackage[caption=false]{subfig}
\usepackage{nicefrac}
\usepackage{graphicx}
\usepackage{cite}  
\usepackage[small]{caption2}
\usepackage{mathtools}
\usepackage{nccfoots}
\usepackage{tikz}
\usepackage{dsfont}
\usepackage{siunitx}
\usepackage[nolist]{acronym}

\usetikzlibrary{calc,positioning}

\usepackage{pgfplots}%
\pgfplotsset{compat = newest}

\newcommand{\Z}[1]{\ensuremath{\mathds{Z}_{#1}}} 

\newcommand{\e}{\mathrm{e}}
\newcommand{\dd}{\mathrm{d}}

\newcommand{\vev}[1]{\ensuremath{\langle{#1}\rangle}}

\usepackage{xcolor}
\definecolor{darkgreen}{HTML}{109930}
\definecolor{pink}{rgb}{0.858, 0.188, 0.478}

\advance \headheight by 3.0truept       
\usepackage[pdftex,hidelinks,colorlinks=true,citecolor=darkgreen]{hyperref}
\hypersetup{
    pdftitle = {A multi-axion model of inflation and dark matter},
    pdfauthor = {Garcia, Gordillo-Ruiz, Henriquez-Ortiz, Ramos-Sanchez}
}
\usepackage{cleveref}
   
\makeatletter
\g@addto@macro\bfseries{\boldmath}
\makeatother

\begin{document}
\begin{acronym}
  \acro{EFT}{effective field theory}
  \acro{ALP}{axion-like particle}
  \acro{BBN}{Big-Bang Nucleosynthesis}
  \acro{DM}{dark matter}
  \acro{DDM}{dynamical dark matter}
  \acro{SD}{spectral distortion}
  \acro{CMB}{cosmic microwave background}
  \acro{KK}{Kaluza-Klein}
  \acro{WIMP}{weakly interacting massive particle}
  \acro{SM}{Standard Model}

  \acro{BSM}{beyond the standard model}
  \acro{BU}{bottom-up}
  \acro{LEET}{low-energy effective theory}
  \acro{QFT}{quantum field theory}
  \acro{RG}{renormalization group}
  \acro{RGE}{renormalization group equation}
  \acro{SUSY}{supersymmetry}
  \acro{TD}{top-down}
  \acro{UV}{ultraviolet}
  \acro{VEV}{vacuum expectation value}
\end{acronym}

\begin{titlepage}

\vspace*{1.0cm}

\begin{center}
{\Large\textbf{\boldmath 
A multi-axion model of inflation and dark matter
}
\unboldmath}

\vspace{1cm}
\textbf{Marcos A.~G.\ Garc\'ia$^\dagger$\footnote{\texttt{marcos.garcia@fisica.unam.mx}}},
\textbf{Hansel Gordillo--Ruiz$^\dagger$\footnote{\texttt{hanselgordillo@ciencias.unam.mx}}},
\textbf{Ra\'ul Henr\'iquez--Ortiz$^*$\footnote{\texttt{raul.henriquez@ues.edu.sv}}},\\
 and
\textbf{Sa\'ul Ramos--S\'anchez$^\dagger$\footnote{\texttt{ramos@fisica.unam.mx}}}
\\[5mm]
\textit{\small $^\dagger$ Instituto de F\'isica, Universidad Nacional Aut\'onoma de M\'exico, Cd.~de M\'exico,\\ C.P.~04510, M\'exico}\\[1mm]
\textit{\small $^*$ Escuela de F\'isica, Facultad de Ciencias Naturales y Matem\'atica, Universidad de El Salvador,\\
final de Av. M\'artires y H\'eroes del 30 julio, San Salvador, C.P.~1101, El Salvador}
\end{center}

\vspace{1cm}

\vspace*{1.0cm}

\begin{abstract}
Models with extra dimensions include a tower of massive states that unavoidably contribute to cosmological dynamics.
In particular, after compactification, a higher-dimensional axion gives rise to a set of states whose dynamics at
early times exhibit the properties of inflation and, at late times, those of dynamical dark matter.
Interestingly, natural choices of parameters imply that the inflationary trajectory in field space is approximately geodesic,
meaning that inflation is effectively driven by a single inflaton corresponding to the heaviest state. In this scenario,
all inflationary observables, including the emerging spectral distortions, are consistent with current observational
constraints, and exact numerical solutions of the equations of motion show that reheating can consistently proceed
down to temperatures of a few MeV. Furthermore, all states lighter than the inflaton only come to dominate the energy density
of the Universe during the matter-dominated era, thus behaving as dark matter at late times.
These states undergo a cascade of decays into radiation, while maintaining a sufficient dark-matter abundance to account for the observed relic density. The lightest dark-matter
candidate remains stable on timescales far exceeding the age of the Universe, thereby satisfying even the most conservative
observational constraints.
\end{abstract}

\end{titlepage}

\newpage


\section{Introduction}  

Modern cosmology is remarkably well described by the $\Lambda$CDM model,  when we include an epoch of rapid accelerated expansion in the early Universe, known as cosmic inflation~\cite{Guth:1980zm,Linde:1981mu}. Besides explaining the observed large-scale homogeneity and isotropy of the Universe, inflation naturally predicts an almost spatially flat geometry, dilutes unwanted relics such as magnetic monopoles, and generates the primordial fluctuations that later evolve into cosmic structures~\cite{Liddle:1999mq,Baumann:2009ds}. This scenario has been tested via high-precision measurements of anisotropies in the \ac{CMB}~\cite{Planck:2018nkj,Hu:2001bc}, which have so far provided compelling evidence of the inflationary paradigm. At the same time, the $\Lambda$CDM framework implies that nearly 85\% of the matter content of the Universe consists of \ac{DM}, whose fundamental nature remains unknown. Hence, understanding the physics responsible for inflation and the nature of \ac{DM} remains among the central challenges of contemporary cosmology.

A wide variety of mechanisms have been proposed to address these two phenomena. In many inflationary models, the accelerated expansion of the early Universe is driven by the dynamics of a scalar or pseudo-scalar field whose potential energy dominates the cosmic energy budget for a short time in the early universe~\cite{Lyth:1998xn,Liddle:2000cg,Pajer:2013fsa}. Likewise, numerous \ac{DM} scenarios postulate the existence of new electrically neutral particles whose properties are consistent with cosmological observations while remaining compatible with current experimental constraints~\cite{Feng:2010gw,Bozorgnia:2024pwk}. From a theoretical perspective, it is particularly appealing to investigate frameworks in which a common sector is responsible for both inflation and \ac{DM}, thereby providing a more economical and perhaps elegant description of the dark universe.

Among the candidates for such a unified framework, \acp{ALP} occupy a privileged position. These pseudo-scalar gauge singlets, closely related to the QCD axion originally introduced to solve the strong CP problem~\cite{Peccei:2006as}, have been shown to arise naturally and play a cosmological role in a broad range of extensions of the \ac{SM}~\cite{Kim:1983dt,Randall:1992ut,Covi:1999ty,Choi:2011yf,Higaki:2016yqk,Calibbi:2016hwq,DiLuzio:2020wdo}, especially in string-inspired constructions~\cite{Svrcek:2006yi,Arvanitaki:2009fg,Choi:2009jt,Acharya:2010zx,Cicoli:2012sz,Halverson:2017deq,Petrossian-Byrne:2025mto,Leedom:2025mlr,Gordillo-Ruiz:2025xcg}. Their weak interactions and protected potentials endow them with a rich cosmological phenomenology, making them suitable candidates for both inflationary dynamics and \ac{DM}. Indeed, axion-like fields play a central role in many successful bottom-up inflationary scenarios~\cite{Freese:1990rb,Adams:1992bn,Kim:2004rp,Dimopoulos:2005ac,Czerny:2014wza,Daido:2017tbr,Abe:2023ylh,Lee:2023dtw,Berghaus:2024zfg,Huang:2025yfo} and numerous string-motivated constructions~\cite{Baumann:2009ni,Cicoli:2011zz,Marchesano:2014mla,Blumenhagen:2014nba}. At the same time, ultralight \acp{ALP}~\cite{Halverson:2017deq,Ferreira:2020fam} are among the most extensively studied \ac{DM} candidates away from the \ac{WIMP} paradigm~\cite{Arcadi:2017kky,Arcadi:2024ukq}.

An additional layer of structure emerges when \acp{ALP} propagate in extra dimensions. Upon compactification, a higher-dimensional \ac{ALP} gives rise to an infinite tower of \ac{KK} excitations, each of which can participate in the cosmological evolution of the Universe. Such towers naturally lead to multifield dynamics at early times and rich dark-sector phenomenology at late times. In particular, they provide the basis for the \ac{DDM} framework~\cite{Dienes:2011ja,Dienes:2011sa}, where the dark sector consists of a large ensemble of unstable \ac{KK} states whose collective abundance reproduces the observed \ac{DM} density through a nontrivial balance between lifetimes and relic abundances. These constructions demonstrate that higher-dimensional axion sectors can generate viable and highly nonstandard cosmological histories.

Motivated by these developments, we build upon the framework known as dark inflaxion~\cite{Gordillo-Ruiz:2023qxi}. In this scenario, a single five-dimensional \ac{ALP} simultaneously drives cosmic inflation and accounts for the \ac{DM} content of the Universe. After compactification, the resulting \ac{KK} tower naturally interpolates between these two epochs: the heaviest mode acts as the inflaton during the early Universe, while the lighter modes survive as a \ac{DM} ensemble at late times. The proposed model therefore provides a unified higher-dimensional realization of inflation and \ac{DM}, linking extant open questions in cosmology to the dynamics of a single underlying field.

This work is organized as follows.
\Cref{sec:model} is devoted to introducing the main elements that constitute our framework.
In \Cref{sec:Inflation}, we discuss the compatibility of our model with a broad set of inflationary constraints.
\Cref{sec:spectral} studies the inflationary consequences for spectral distortions of our model and their detectability.
In \Cref{sec:DM}, we address the dynamics of \ac{DM} emerging from our model and show its compatibility with observations.
Our conclusions and outlook are provided in \Cref{sec:conclusions}.

\section{Dark inflaxion: unifying inflation and dark matter}
\label{sec:model}

Dark inflaxion~\cite{Gordillo-Ruiz:2023qxi} is inspired by the framework of \ac{DDM}~\cite{Dienes:2011ja,Dienes:2011sa} together with higher-dimensional realizations of hilltop inflation~\cite{Daido:2017tbr} in top-down constructions, endowed with extra dimensions. In particular, we consider a five-dimensional axion-like sector compactified on an orbifold geometry, where the collective dynamics of \ac{KK} modes simultaneously account for inflationary dynamics and \ac{DM} phenomenology.

Theories with extra spatial dimensions naturally lead, after compactification, to an effective four-dimensional description containing an infinite tower of \ac{KK} states. The properties of these states are determined by the compactification geometry and by the boundary conditions imposed on the higher-dimensional fields. Experimental and astrophysical observations constrain the compactification radius through deviations from Newtonian gravity, supernova cooling bounds, and neutron-star heating constraints~\cite{Workman:2022ynf}. Representative upper bounds for different numbers of extra dimensions are summarized in \Cref{table:limitsRadius}.

We consider a single extra spatial dimension compactified on the orbifold $S^1/\Z2$, such that the full spacetime geometry is $\mathcal{M}_5 = \mathcal{M}_4 \times \left(S^1/\Z2\right)$, and $x^M=(x^\mu, y)^\mathrm{T}$ denotes the five-dimensional coordinates, where $x^\mu$ with $\mu=0,1,2,3$ correspond to the ordinary four-dimensional Minkowski spacetime, while $y$ parametrizes the compact extra dimension. The orbifold construction is obtained by imposing the identifications
\begin{equation}
y~\sim~ y+2\pi R\,,
\qquad
y~\sim~ -y\,,
\end{equation}
where $R$ is the compactification radius. Consequently, a bulk scalar field must satisfy
\begin{equation}
\Phi(x^\mu,y)~\stackrel!=~\Phi(x^\mu,y+2\pi R)\,,
\qquad
\Phi(x^\mu,y)~\stackrel!=~\Phi(x^\mu,-y)\,.
\end{equation}
Under these conditions, the scalar field $\Phi$ can be expressed in terms of a \ac{KK} tower as~\cite{Dienes:1999gw}
\begin{align}
\label{eq:expansion}
	\Phi(x^\mu,y) ~=~ \frac{1}{\sqrt{2\pi R}}\sum^{\infty}_{m=0}\,r_m
	\phi^m(x^\mu) \cos\left( \frac{m\, y}{R} \right)\,,
\end{align}
where the normalization coefficients are  $r_0=1$ and $r_m=\sqrt{2}$ for $m>0$, ensuring canonical normalization of the four-dimensional fields $\phi^m(x^\mu)$.

\begin{table}[t]
	\centering
	\begin{tabular}{|c | c |  c |  c|}
		\hline
		$D$ & 1 & 2 & 3 \\
		\hline\hline
		Torsion-balance & $\num{2.22e11}$\,GeV$^{-1}$ & --                         & -- \\   \hline
		SN 1987A        & $\num{2.48e18}$\,GeV$^{-1}$ & $\num{4.86e9}$\,GeV$^{-1}$ & $\num{5.77e6}$\,GeV$^{-1}$\\ \hline
		Neutron-star excess heat & $\num{2.25e11}$\,GeV$^{-1}$ & $\num{7.85e5}$\,GeV$^{-1}$& $\num{1.30e4}$\,GeV$^{-1}$\\ [.5ex]
		\hline
	\end{tabular}
	\caption{Upper limits on the compactification radius of $D=1,2,3$ extra dimensions, determined
		from torsion-balance experiments~\cite{Kapner:2006si}, 1987A supernova observations and
		neutron-star excess heat measurements~\cite{Hannestad:2003yd}.}
	\label{table:limitsRadius}
\end{table}

We consider $\Phi$ to be a massless five-dimensional axion-like field. Observable \ac{SM} fields are assumed to be confined to the orbifold fixed point at $y=0$, while the axion propagates in the full five-dimensional spacetime. Consequently, interactions between the axion sector and ordinary matter occur only on the brane. Assuming only electromagnetic interactions, the five-dimensional Lagrangian is therefore given by
\begin{align}
	\mathcal{L}_{5\text{-}\mathrm{d}} ~\supset~
	\frac{1}{2}\partial^M \Phi \partial_M \Phi-
	\delta(y)\frac{g_{\Phi\gamma\gamma}}{4}\,\Phi \,F_{\mu\nu}\widetilde{F}^{\mu\nu}\,,
\end{align}
where the coupling $g_{\Phi\gamma\gamma}$ has mass dimension $\nicefrac{-3}{2}$. Substituting the \ac{KK} decomposition of \cref{eq:expansion} into the five-dimensional Lagrangian and integrating over the compact dimension yields the effective four-dimensional action
\begin{align}
\label{eq:4-d_action}
	S~\supset~\int \dd^4x\; \left(\frac{1}{2}\sum^{\infty}_{m=0}\sum^{\infty}_{n=0}g_{mn}\partial^\mu
	\phi^m \partial_\mu \phi^n -\frac{1}{2}\sum^{\infty}_{m=0}\frac{m^2}
	{R^2}(\phi^m)^2 -\frac{g_{a\gamma\gamma}}{4}\sum^{\infty}_{m=0} r_m \phi^m
	F_{\mu\nu}\widetilde{F}^{\mu\nu}\right)\,,
\end{align}
where the field-space metric is $g_{mn}=\delta_{mn}$.
The four-dimensional axion coupling constant is defined as $g_{a\gamma\gamma}:=g_{\Phi\gamma\gamma}/\sqrt{2\pi R}$. Another essential ingredient of the model is a non-perturbatively generated brane-localized potential characterized by the scale $\Lambda$. We consider a five-dimensional hilltop-like axion potential of the form
\begin{equation}\label{eq:Vinstanton}
	V_\mathrm{5-d} ~\supset~ \delta(y)\,\Lambda^4\left[\cos \left(\frac{\Phi}{f^{3/2}_{\Phi}}
	+ \Theta \right) -\frac{\kappa}{n^2}\cos\left(\frac{n\,\Phi}{f^{3/2}_{\Phi}}\right)\right]\,,
\end{equation}
where $f_\Phi$ is the fundamental five-dimensional decay constant and $\Theta$ denotes a phase. After dimensional reduction, the effective four-dimensional potential becomes
\begin{align}
\label{eq:4-d_potential}
	V ~\supset~ \Lambda^4\left[
	\cos\left(\frac{\sum^{\infty}_{m=0}r_m  \phi^m}{f_a}+\Theta\right)
	-\frac{\kappa}{n^2}\cos\left(\frac{n\sum^{\infty}_{m=0}r_m  \phi^m}{f_a} \right)
	\right]\,,
\end{align}
where the effective four-dimensional decay constant is  $f_a:= \sqrt{2\pi R} f^{3/2}_\Phi$. Since the effective four-dimensional theory originates from a single five-dimensional axion field, all \ac{KK} modes share the same decay constant $f_a$. This differs from conventional bottom-up multi-axion constructions, where independent decay constants are typically associated with different fields.

Although the \ac{KK} tower formally contains infinitely many states, only a finite subset remains
dynamically relevant below an effective {\it cutoff scale}. Assuming that only $M+1$ modes remain
light, the potential becomes
\begin{align}
\label{eq:4-d_potentialN}
	V~=~\sum^{M}_{m=0}\frac{m^2}{2R^2}(\phi^m)^2+\Lambda^4\left[
	\cos\left( \frac{\sum^{M}_{m=0}r_m\phi^m}{f_a} +\Theta\right) -\frac{\kappa}{n^2}
	\cos\left( \frac{n\sum^{M}_{m=0}r_m\phi^m}{f_a} \right)\right].
\end{align}
Minimizing the potential yields
\begin{equation}\label{eq:minphi0}
\vev{\phi_\mathrm{min}^0}~=~
\left(
\pi \ell
-
\left[
\frac{6\Theta}{\kappa n^2-1}
\right]^{1/3}
\right)f_a,
\qquad
\ell\in 2\Z{}+1,
\end{equation}
and
\begin{equation}\label{eq:minphim}
   \vev{\phi_\mathrm{min}^m} ~=~ 0\,, \qquad m>0\,.
\end{equation}
We now study the properties of the physical four-dimensional mass eigenstates. Following the
framework of \ac{DDM}, the mass eigenstates are denoted by $\pmb{a}=(a^0, a^1, \ldots, a^M)^{\text{T}}$, and are related to the \ac{KK} basis through
\begin{align}
\label{eq:Ppsi}
	\pmb{a}~=~R\,\pmb{\phi}\,.
\end{align}
Here, $R$ denotes the matrix that diagonalizes the squared mass matrix $\mathcal{M}^2$, whose components are defined by
\begin{align}
	\mathcal{M}^2_{m m'} &:=\frac{\partial^2 V(\pmb{\phi})}{\partial \phi^m \partial \phi^{m'}}\\
	&=\frac{m^2}{R^2}\delta_{m m'}+
	\frac{\Lambda^4}{f^2_a}r_mr_{m'}\left[ \kappa \cos\left( \frac{n \sum^{M}_{m''=0}r_{m''} \phi^{m''}}{f_a} \right) -\cos\left( \frac{\sum^{M}_{m''=0}r_{m''} \phi^{m''}}{f_a}+\Theta \right)\right]\,.\nonumber
\end{align}
Using eq.~\eqref{eq:4-d_potentialN}, the squared mass matrix becomes
\begin{align}
\label{eq:massmatrix}
	\mathcal{M}^2_{m m'}&~\simeq~\frac{m^2}{R^2}\delta_{m m'}+\frac{\Lambda^4}{f^2_a}r_m r_{m'}\left[ 1+\kappa \cos(n \pi) - \frac12\left(\frac{6 \Theta}{\kappa\, n^2-1}\right)^{\nicefrac23}\left(1+\kappa \, n^2 \cos (n\pi)\right)\right] \notag    \\
	&~\simeq~M^2_c m^2\delta_{m m'}+m^2_\Lambda\,r_m r_{m'} \,.
\end{align}
In the last equation, we have adopted the benchmark values $n=3$ and $\kappa = 1$. The
compactification scale $M_c$ and the natural scale of the axion $m_\Lambda$ are introduced and defined here as
\begin{align}
	\label{eq:m_lambda}
	M^2_c         ~:=~ \frac{1}{R^2} \qquad \text{and}\qquad
	m^2_{\Lambda} ~:=~ (6\Theta)^{\nicefrac23}\frac{\Lambda^4}{f^2_a}\,.
\end{align}
The interplay between these two scales controls the inflationary dynamics and the properties of
the \ac{DM} ensemble. In the following, we shall take $m_\Lambda>M_c$ to be the cutoff scale of the model.

As we shall explicitly show in the following sections, the interplay between the compactification scale $M_c$, the non-perturbative scale $\Lambda$, and the effective decay constant $f_a$ simultaneously determines: the inflationary observables, the reheating dynamics, the spectral distortion signal, and the late-time \ac{DM} abundances. Therefore, the model provides a unified higher-dimensional realization in which inflation, reheating and \ac{DM} production emerge from the collective dynamics of a single five-dimensional axion sector.

\section{Inflation}
\label{sec:Inflation}

To ensure the viability of our scenario for inflation, we impose two basic cosmological 
requirements: a reheating temperature above the lower bound imposed by \ac{BBN} and a 
sufficiently long-lived lightest eigenstate to account for the \ac{DM} abundance. As we 
shall shortly see, these requirements provide a simple physical criterion for selecting 
the hierarchy between the cutoff scale, which is taken to be $m_\Lambda$,
and the compactification scale $M_c$, as well as the number of axion fields in 
the ensemble.

The reheating temperature $T_\mathrm{RH}$ can be estimated by assuming that the Universe 
is dominated by the heaviest eigenstate prior to its decay and that its decay products 
thermalize rapidly. As will be shown explicitly in our numerical example, the former 
assumption is indeed satisfied in the region of parameter space considered.
The reheating temperature can be estimated by the condition
$\Gamma_M \simeq \tfrac{3}{2}H$, which yields
\begin{equation}
   T_\mathrm{RH} ~\simeq~ \left(\frac{40}{\pi^2 g_\mathrm{RH}} \right)\sqrt{\Gamma_M M_\mathrm{Pl}}\,,
\end{equation}
where $\Gamma_M$ is the decay rate of the heaviest eigenstate $a^M$.
To preserve the successful predictions of \ac{BBN}~\cite{Fields:2019pfx, Hasegawa:2019jsa}, the reheating temperature
must satisfy $T_\mathrm{RH}\gtrsim \mathcal{O}(1\, \mathrm{MeV})$.
At the same time, the largest contribution to the \ac{DM} abundance is always
provided by the lightest eigenstate $a^0$, which is required to be sufficiently long-lived to be
compatible with indirect searches for decaying \ac{DM}~\cite{Cohen:2016uyg, Blanco:2018esa}. Throughout this
work, we adopt the conservative requirement
\begin{equation}
	\tau_0 ~=~ \frac{1}{\Gamma_0} ~\gtrsim~ 10^{25} \, \mathrm{s} \, ,
\end{equation}
which is weaker than the strongest current observational bounds
for most masses and decay channels.
By setting $T_\mathrm{RH}\simeq 4\,\mathrm{MeV}$, which lies above the lower
bound from \ac{BBN}, we obtain the ratio
between the decay rates of the heaviest and the lightest states
\begin{equation}\label{eq:RHDMquot}
	\frac{\Gamma_M}{\Gamma_0} ~=~ \frac{\num{1.07e-23}\, \mathrm{GeV}}{\num{1.52e-49} \, \mathrm{GeV}} ~\gtrsim~ \num{1.63e26} \, . 
\end{equation}
Here, as an illustrative example, we set the cutoff scale in
\cref{eq:4-d_potentialN} to $m_\Lambda \simeq 300\,M_c$, which corresponds to
an ensemble of $M+1=300$ fields. This choice yields a ratio
$\Gamma_{299}/\Gamma_0 \simeq \num{3e26}$, corresponding to a lifetime
$\tau_0 = 10^{25}\,\mathrm{s}$ and a reheating temperature of
$T_\mathrm{RH}\approx5.42\,\mathrm{MeV}$. Increasing the cutoff scale further enhances 
the hierarchy between the decay rates, making it easier to achieve both a higher 
reheating temperature and a sufficiently long-lived \ac{DM} candidate. The full set of 
parameter values\footnote{The origin of a trans-Planckian value of $f_a$ has been long discussed in the literature, where some possibilities have been identified in the presence of many axions~\cite{Kim:2004rp,Burgess:2014oma,Kim:2017tdk}.} 
for the benchmark example is given in \Cref{tab:bestfit}.

\begin{table}[tb]
\centering
\begin{tabular}{ll|ll}
	\toprule
	\multicolumn{2}{c}{Model parameters} & \multicolumn{2}{c}{Resulting quantities} \\
	\midrule
	$M_c$ & $\num{1.57e6}\,\mathrm{GeV}$ &
	$m_0$ & $\num{7.86e5}\,\mathrm{GeV}$ \\
	
	$\Lambda$ & $\num{3.04e14}\,\mathrm{GeV}$ &
	$m_{299}$ & $\num{1.15e10}\,\mathrm{GeV}$ \\
	
	$f_a$ & $\num{6.92e18}\,\mathrm{GeV}$ &
	$a_{\rm min}^{299}$ & $-\num{8.83e17}\,\mathrm{GeV}$ \\
	
	$\kappa$ & $1$ &
	$\lambda_{299}$ & $4.45$ \\
	
	$n$ & $3$ &
	$g_{a\gamma\gamma}$ & $\num{2.08e-27}\,\mathrm{GeV}^{-1}$ \\
	
	$\Theta$ & $\num{7.30e-6}$ &
	& \\
	
	$\Psi$ & $-3.1$ &
	& \\
\bottomrule
\end{tabular}
\caption{\label{tab:bestfit}
		Parameter values for a benchmark example, yielding a small value of
		$\chi_{n_s}^2+\chi_{A_s}^2+\chi_{\alpha_s}^2$ with respect to the inflationary
		observables given in \cref{eq:Plancknsrvalues,eq:PlanckAsdnsvalues}. This benchmark is also
		compatible with the \ac{BBN} reheating bound and a sufficiently long-lived
		\ac{DM} candidate.}
\end{table}

Having specified the benchmark parameters, we can now investigate the
cosmological evolution of the axion ensemble and compute the corresponding
inflationary observables. At early times, when the decay rate satisfies $\Gamma \ll H$,
the system evolves according to the background equations of motion,

\begin{subequations}\label{subeq:eoms}
	\begin{align}
	H^2  &~:=~ \left(\frac{\dot{\mathrm{a}}}{\mathrm a}\right)^2 ~=~ \frac{1}{3 M_\mathrm{Pl}^2} \left(\frac{1}{2}\dot{\varphi}^2 + V \right) \, , \\
	0 &~=~ \ddot{\phi}^i + 3 H \dot{\phi}^i
	+ g^{ij}\,V_j \, ,
	\label{eq:motioneq}
	\end{align}
\end{subequations}
where $\dot{\varphi}^2 := g_{ij}\dot{\phi}^i\dot{\phi}^j$, $V_i := \partial V / \partial \phi^i$
and the Christoffel symbols have been omitted because the metric in field space is trivial.
We now introduce the slow-roll parameters that characterize a sustained period 
of inflation. The first slow-roll parameter follows from requiring that the 
fractional variation of the Hubble parameter per e-fold be small, namely
\begin{equation}\label{eq:epsilon_parameter}
    \epsilon ~:=~ -\frac{\dot{H}}{H^2} ~=~ \frac{\dot{\varphi}^2}{2 M_\mathrm{Pl}^2 H^2} ~\ll~ 1 \, .
\end{equation}
This condition ensures a quasi-exponential expansion of the
scale factor $\mathrm{a}(t)$. In order to solve the horizon problem, this accelerated phase
must persist for a sufficiently long time. This, in turn, requires that 
$\epsilon$ remains small over many Hubble times. The second slow-roll parameter 
quantifies the stability of this condition,
\begin{equation}\label{eq:eta_parameter}
   \eta ~:=~ \frac{\dot{\epsilon}}{\epsilon H}	~\ll~ 1 \, .
\end{equation}
When both conditions in \cref{eq:epsilon_parameter,eq:eta_parameter} are satisfied, 
the system is said to be in the slow-roll regime.
We define the number of e-folds between horizon crossing and the end of inflation,
which is given by
\begin{equation}\label{eq:efolds}
	N_\star ~=~ \int_{t_\star}^{t_\mathrm{end}}H \dd\tau \, ,
\end{equation}
where $t_\star$ denotes the time at which the pivot scale exits the 
horizon, and $t_\mathrm{end}$ is defined by the condition $\epsilon = 1$, 
which signals the end of inflation. 
Throughout this work we adopt the pivot scale 
$k_\star = 0.05\,\mathrm{Mpc}^{-1}$, corresponding to the choice 
used in the Planck analysis~\cite{Planck:2018vyg}.

The e-fold number at horizon crossing is related to the
post-inflationary evolution via~\cite{Liddle:2003as, Martin:2010kz}
\begin{align}
\label{eq:Totalefolds}
N_\star&~=~\ln\left[\frac{1}{\sqrt{3}}\left(\frac{\pi^2}{30} \right)\left(\frac{43}{11} \right)^{\!1/3} \frac{T_0}{H_0} \right]
-\ln\left( \frac{k_\star}{\mathrm{a}_0H_0}\right)-\frac{1}{12}\ln g_\mathrm{RH} \nonumber\\
&~+~\frac{1}{4}\ln \left(\frac{V_\star^2}{M_\mathrm{Pl}^4 \rho_\mathrm{end}}\right)
+ \frac{1-3 w_\mathrm{int}}{12(1+w_\mathrm{int})}\ln\left(\frac{\rho_\mathrm{RH}}{\rho_\mathrm{end}} \right) \, ,
\end{align}
where $V_\star := V(\pmb{\phi}_\star)$ is the potential evaluated at horizon crossing.
We set the present scale factor to $\mathrm{a}_0=1$, take the Hubble parameter today as 
$H_0=67.66 \, \mathrm{km}\,\mathrm{s}^{-1}\,\mathrm{Mpc}^{-1}$~\cite{Planck:2018vyg},
and the CMB temperature as $T_0=2.7255 \,\mathrm{K}$~\cite{Fixsen_2009}.
Here, $\rho_\mathrm{end}$ denotes the energy density at the end of inflation,
$\rho_\mathrm{RH}$ is the energy density at the end of reheating,
$w_\mathrm{int}$ is the effective (e-fold averaged) equation-of-state parameter
during reheating, and $g_\mathrm{RH}=10.75$ is the effective number of
relativistic degrees of freedom at that epoch, counting them in the Standard Model
for a low reheating temperature (see \Cref{sec:reheating}).

\begin{figure}[t]
\centering
\subfloat{%
		\includegraphics[width=0.49\textwidth]{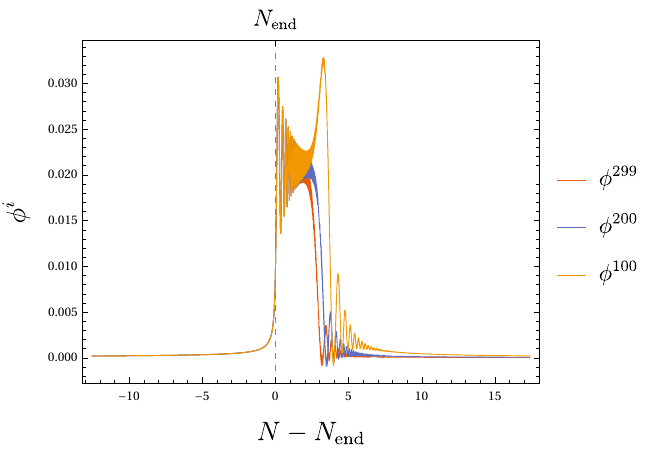}
}
\hspace{0.25cm}
\subfloat{%
		\includegraphics[width=0.40\textwidth]{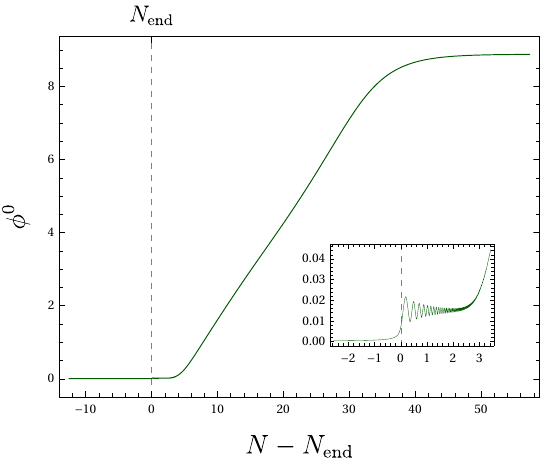}
}
\caption{\label{fig:dynamics}
		Evolution of the fields $\phi^i$ as a function of the number of e-folds $N$ (the superscript is not
		the power of $\phi$).
		The trajectories were obtained by numerically solving the equations of motion~\eqref{subeq:eoms} for
		$300$ fields with the parameters shown in \Cref{tab:bestfit}. 
		The left panel displays the evolution of a representative subset of fields, 
		while the right panel shows the evolution of the field $\phi^0$. 
		The vertical line $N_{\mathrm{end}}$ denotes the end of inflation, defined by the condition $\epsilon = 1$.
	}
\end{figure}

We solve \cref{subeq:eoms} numerically using the potential defined in
\cref{eq:4-d_potentialN} for our $300$ fields, i.e.\ with $M=299$. 
We use the parameters given in \Cref{tab:bestfit}.
For numerical convenience, we adopt the number of e-folds $N$, defined through 
$\dd N = H\,\dd t$, as the time variable.
Although one would generically expect a genuinely multifield inflationary
dynamics, we will show below that the evolution effectively reduces to that 
of a single-field system. 
For the time being, however, we treat the dynamics in the full multifield framework.
The dynamics of the fields can be seen in \Cref{fig:dynamics}.

We observe that the fields exhibit two distinct oscillatory regimes. 
The first is driven by the quartic contribution to the potential~\eqref{eq:4-d_potential}, 
while the second arises once the quadratic term becomes dominant. 
During the inflationary phase we find $\eta \gg \epsilon$, as typically expected in
small-field models of inflation.
Although inflation is often defined to end when one of the two slow-roll parameters is $ 1$, 
in our numerical analysis we instead adopt the condition 
$\ddot{\mathrm{a}} = 0$ (corresponding always to $\epsilon=1$) to mark the end of inflation.
With this definition, the total duration of inflation is  around $70$ e-folds, sufficient to solve the horizon problem.

\begin{figure}[t]
	\centering
	\subfloat{%
		\includegraphics[width=0.49\textwidth]{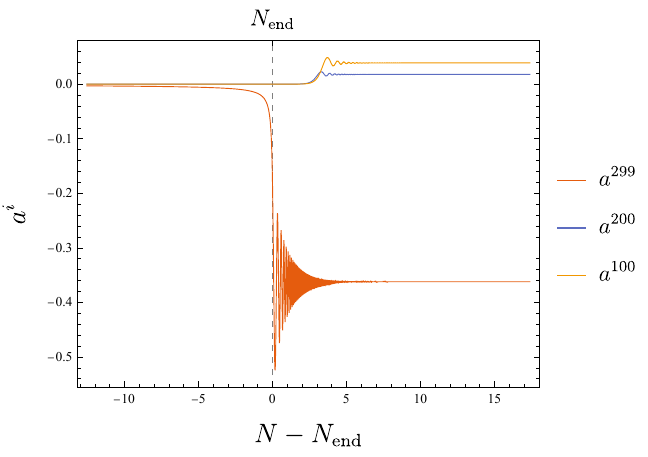}
	}
	\hspace{0.25cm}
	\subfloat{%
		\includegraphics[width=0.40\textwidth]{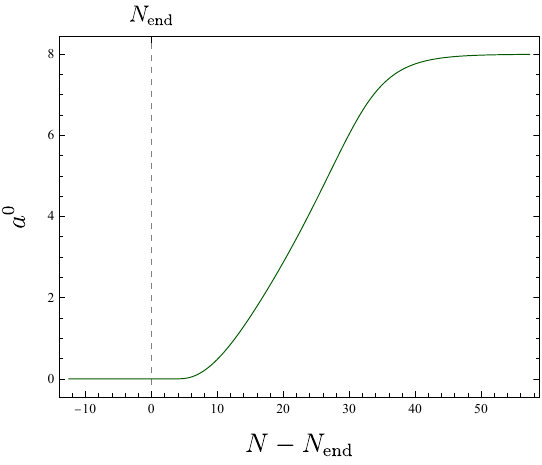}
	}  
	\caption{\label{fig:dynamicseigenstates}
		Evolution of the mass eigenstates $a^i$ as a function of the number of e-folds
		$N$. The left panel illustrates the evolution of a representative subset of the
		heavier eigenstates, while the right panel focuses on the lightest eigenstate,
		which later constitutes the dominant \ac{DM} component. The evolution is
		obtained from the numerical solution of the background equations of
		motion~\eqref{subeq:eoms} using the benchmark parameters listed in
		\Cref{tab:bestfit}. The vertical line identifies the end of inflation, as set by
		$\epsilon=1$.
	}
\end{figure}

It is also important to solve the equations of motion in terms of 
the mass eigenstates $a^i$, since this description becomes 
particularly relevant during subsequent epochs, such as reheating. 
To this end, we diagonalize the mass matrix in \cref{eq:massmatrix} and obtain the corresponding eigenstates. The mass eigenstates are related to the original field basis through
\begin{equation}
   a^i ~=~ \sum_{j=1}^{M+1} R^i_j\phi^j \, .
\end{equation}
We then solve the system of equations~\eqref{subeq:eoms} expressed 
in the mass-eigenstate basis $a^i$. The lightest eigenstate has a mass
$m_{0} = \num{7.86e6} \,\mathrm{GeV}$,
whereas the heaviest eigenstate has a mass 
$m_{299} = \num{1.15e10} \,\mathrm{GeV}$.
The resulting dynamics in terms of these eigenstates is shown in \Cref{fig:dynamicseigenstates}.

\subsection{Single field vs multifield}
\label{sec:singmulti}

In the presence of a large number of scalar fields, it is instructive to 
examine whether the dynamics display genuinely multifield behavior,
potentially characterized by significant bending of the background 
trajectory in field space. In order to make this geometric structure 
explicit, we introduce a kinematical basis adapted to the background 
trajectory, which allows us to decompose the motion into tangential 
and transverse directions~\cite{Gordon:2000hv, Peterson:2011yt}.

For a system with more than two scalar fields, this decomposition 
consists of a single tangent vector $T^i$ and an $M$-dimensional normal 
subspace, spanned by orthonormal vectors $N^i_{(I)}$ with 
$I=1,\dots,M$. The unit tangent vector is defined as
\begin{equation}
	T^i ~:=~ \frac{\dot{a}^i}{\dot{\varphi}} \, ,
\end{equation}
where  $\dot{\varphi} := \sqrt{g_{ij}\dot{a}^i\dot{a}^j} $ is the magnitude of
the velocity in field space. Note that we conveniently construct this basis from the mass eigenstates $a^i$. 
The basis vectors satisfy the orthonormality and completeness relations
\begin{equation}
	g_{ij} T^i T^j ~=~ 1 \, ,
	\qquad
	g_{ij} N^i_{(I)} T^j ~=~ 0 \, ,
	\qquad
	g_{ij} N^i_{(I)} N^j_{(J)} ~=~ \delta_{IJ} \, ,
\end{equation}
which allows any field-space vector to be decomposed into 
adiabatic (tangent) and entropic (normal) components. 
In this geometrical framework, purely geodesic motion corresponds to
\begin{equation}\label{eq:geodesic_condition}
	D_t T^i ~=~ \partial_t T^i ~=~ 0 \, ,
\end{equation}
where the covariant time derivative reduces to the time derivative because the metric is trivial in field space.
The condition in \cref{eq:geodesic_condition} then can be written as
a geodesic equation in field space with affine 
parameter proportional to cosmic time. Therefore, any non-vanishing 
projection of $D_t T^i$ onto the normal directions signals a departure 
from geodesic motion and encodes the curvature-induced bending of 
the inflationary trajectory.
It is convenient to parametrize this bending in terms of the 
dimensionless turning rates $\omega_I$, defined through
\begin{equation}
	D_N T^i	= \omega_I N^i_{(I)} \, ,
\end{equation}
where $D_N := H^{-1} D_t$ denotes the covariant derivative with 
respect to the number of e-folds. Notice that there is, in general, 
one independent turning rate $\omega_I$ for each normal direction.
The magnitude of the bending is then quantified by
\begin{equation}
	|D_N T| = \sqrt{\delta^{IJ}\omega_I \omega_J} \, .
\end{equation}
If $|D_N T| \ll 1$, the trajectory follows an approximately geodesic 
path in field space and the dynamics effectively reduce to a 
single-field description. Conversely, sizable values of 
$|D_N T|$ signal genuinely multifield effects and the possible
excitation of entropic fluctuations.

\begin{figure}
	\centering
	\includegraphics[width=0.42\linewidth]{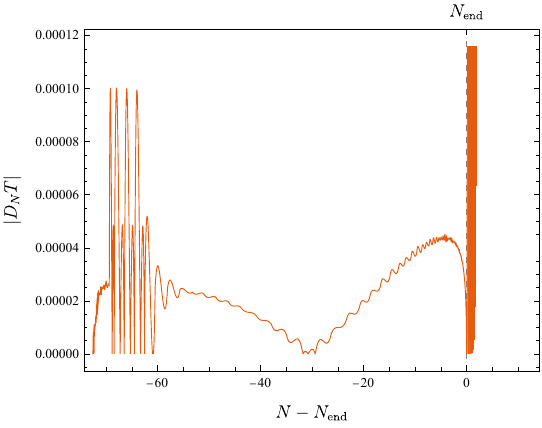}
	\caption{Evolution of $|D_N T|$ during inflation. The end of inflation occurs at $N_{\rm end}=72.57$. We find $|D_N T| \simeq 0$ throughout the inflationary phase, indicating that the background trajectory follows an (approximately) geodesic motion.}
	\label{fig:turning-rate}
\end{figure}

With this in mind, we evaluate the quantity $|D_N T|$ along the inflationary trajectory described above. 
The result is presented in \Cref{fig:turning-rate}. We find that the magnitude of the turning rate 
remains small, $|\omega_I| \ll 1$, throughout the entire evolution. As a consequence, the total bending
during inflation is negligible. This indicates that the background trajectory is well approximated by
a geodesic in field space, and that multifield effects associated with significant turning are strongly
suppressed. In this regime, the dynamics are effectively captured by a single adiabatic degree of freedom.

\begin{figure}
	\centering
	\includegraphics[width=0.6\linewidth]{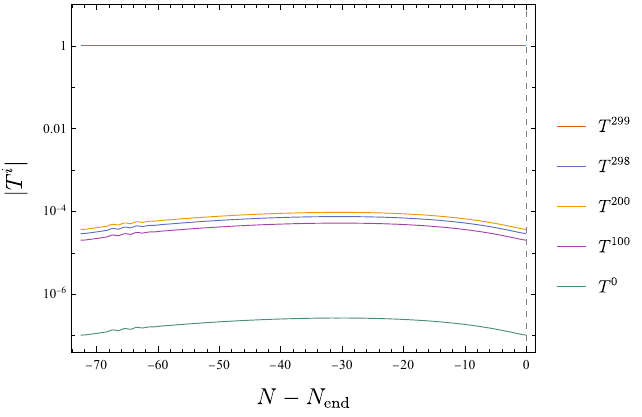}
	\caption{Evolution of the components of the tangent vector $T^i$ during the inflationary epoch, expressed in the eigenstate basis $a^i$. We observe that the component along $a^{299}$ clearly dominates, indicating that this field provides the only dynamical degree of freedom throughout inflation.}    
	\label{fig:tangent}
\end{figure}

We can now identify an important property of the model by analyzing 
the components of the tangent vector $T^i$. Since $T^i$ determines 
the instantaneous direction of motion in field space, its components 
quantify how the inflationary trajectory is distributed among the 
different mass eigenstates. As shown in \Cref{fig:tangent}, the dominant contribution to 
$T^i$ arises from the heaviest eigenstate, $a^{299}$. This indicates 
that the background evolution is overwhelmingly aligned along this 
direction in field space, while the remaining fields provide only 
subleading contributions. It is therefore natural to identify 
$a^{299}$ as the effective inflaton of the model.

We conclude that the inflationary dynamics are not only effectively 
single-field due to the negligible turning rate, but also dynamically 
dominated by the heaviest mass eigenstate, with the lighter fields 
remaining essentially frozen along the trajectory.

\subsection{Inflationary observables}

If the inflationary trajectory in field space is geodesic, the system effectively
behaves as an adiabatic single-field model. In this regime a slow-roll approximation
can be implemented, enabling analytical estimates of the CMB observables in terms
of the standard slow-roll parameters~\cite{Lalak:2007vi}. 
If the trajectory were not geodesic, however, isocurvature perturbations could source the
adiabatic mode after horizon crossing, and the single-field expressions would no longer
be valid. In that case the full system of perturbation equations  must be solved
numerically in order to determine the power spectra and the corresponding amplitudes and
spectral tilts.
Fortunately, in our model the inflationary trajectory is geodesic in field space.
Therefore the standard single-field slow-roll expressions for the CMB observables
remain applicable, and we use these analytical formulas in the examples discussed below.

We therefore employ the standard expressions for the cosmological observables
in terms of the slow-roll parameters defined in
\cref{eq:epsilon_parameter,eq:eta_parameter}.
In the slow-roll approximation, the scalar spectral index and the
tensor-to-scalar ratio are given by
\begin{equation}
	n_s ~=~ 1 - 2\epsilon - \eta \,,
	\qquad 
	r ~=~ 16\,\epsilon \, .
\end{equation}
According to the latest results from Planck and BICEP/Keck 2018,
the observed values of the scalar spectral index and the tensor-to-scalar
ratio are
\begin{subequations}\label{eq:Plancknsrvalues}
\begin{align} n_s &~=~ 0.9665\pm 0.0038, &&\text{({\small $68\%$, TT,TE,EE+lowE+lensing+BAO,~\cite[Table 2]{Planck:2018vyg}})}\, , \label{eq:n_s} \\
r &~<~ 0.066,&&({\small 95\%\text{, TT,TE,EE+lowE+lensing+BK15+BAO,~\cite[eq.~(45a)]{Planck:2018vyg}}})\label{eq:r}\, .
\end{align} 
\end{subequations} 
Here $n_s$ is quoted at the pivot scale $k_\star = 0.05\,\mathrm{Mpc}^{-1}$,
while $r$ is given at $k_\star = 0.002\,\mathrm{Mpc}^{-1}$.
The running of the scalar spectral index can likewise be expressed
in terms of the slow-roll parameters,
\begin{equation}\label{eq:running}
\alpha_s ~:=~ \frac{\dd n_s}{\dd\ln k}
~=~ \frac{1}{1-\epsilon}\frac{\dd n_s}{\dd N} \, .
\end{equation}
The amplitude of scalar perturbations is given by
\begin{equation}\label{eq:As}
A_s ~=~ \frac{H^2}{8\pi^2 M_\mathrm{Pl}^2 \epsilon}\, .
\end{equation}
The values for the running of the scalar spectral index and the amplitude
of scalar perturbations reported by Planck are
\begin{subequations}\label{eq:PlanckAsdnsvalues} 
\begin{align} \alpha_s &= -0.006\pm 0.013 &&({\small 95\%\text{, TT,TE,EE+lowE+lensing+BK15+BAO,~\cite[eq.~(45b)]{Planck:2018vyg}}})\, , \label{eq:dnsdlnk}\\
 10^9 A_s &= 2.105 \pm 0.030 &&({\small 68\%\text{, TT,TE,EE+lowE+lensing+ BAO,~\cite[Table~2]{Planck:2018vyg}}})\, . \label{eq:As}
\end{align} 
\end{subequations}
With these expressions, we can compare the cosmological parameters predicted
by our model (taking $N_\star=40.6$ corresponding to a reheating scale $T_\mathrm{RH}\simeq 5 \, \text{MeV}$, see \Cref{sec:reheating})
with the observational constraints summarized in
\Cref{tab:observablescomparison}. We find that the predictions of the
model are consistent with the latest observations reported by Planck,
indicating that our setup provides a viable realization of inflation.

An interesting question is whether our model can also accommodate the recent cosmological constraints reported by ACT~\cite{AtacamaCosmologyTelescope:2025blo}. In particular, the combined analysis including ACT and Planck CMB lensing data together with DESI Y1 baryon acoustic oscillation measurements (P-ACT-LB) favors a slightly larger scalar spectral index, $n_s = 0.9743 \pm 0.0034$. We find that our setup can readily reproduce this value through a small shift of the relative phase, namely $\Theta = \num{8.21e-6}$, while maintaining a tensor-to-scalar ratio that remains negligibly small, $r\simeq \mathcal{O}(10^{-8})$, and therefore fully consistent with observational bounds.

\begin{table}[tb]
	\centering
	\begin{tabular}{lcc}
		\toprule
		Observable & Computed value & Observational value\\
		\midrule
		
		$a_{\mathrm{ini}}^{299}$ & $-\num{2.84e15}\,\mathrm{GeV}$ & --\\
		$a_{\mathrm{end}}^{299}$ & $-\num{4.01e17}\,\mathrm{GeV}$ & --\\
		$N_\star$ & $40.6$ & --\\
		
		\midrule
		
		$n_s$ & $0.9662$ & $0.9665\pm0.0038$\\
		$r$ & $\num{1.386e-8}$ & $<0.106$\\
		$10^9A_s$ & $2.105$ & $2.105\pm0.030$\\
		$\alpha_s$ & $-0.0106$ & $-0.006\pm0.013$\\
		
		\midrule
		
		$\chi_{n_s}^2+\chi_{A_s}^2$
		& $0.0294$
		& --\\
		
		$\chi_{n_s}^2+\chi_{A_s}^2+\chi_{\alpha_s}^2$
		& $0.522$
		& --\\
		
		\bottomrule
	\end{tabular}
		\caption{Comparison between the benchmark values obtained in our model and the
		corresponding observational constraints from
		\cref{eq:Plancknsrvalues,eq:PlanckAsdnsvalues}.}
	\label{tab:observablescomparison}
\end{table}

As discussed above, the dynamics of our setup can be effectively reduced
to a single-field description in which the inflaton is identified with
the field $a^{299}$. It is therefore instructive to construct the
corresponding effective potential for this field. To this end, we fix
the remaining relevant fields to their initial values, obtaining the
single-field potential
\begin{align}\label{eq:V299}
    V(a^{299})&\simeq \sum_{m=0}^{M} \frac{1}{2}m^2 M_c^2 \left((R^{-1})^m_{299}a^{299}\right)^2 \nonumber\\
    &+ \Lambda^4\left[ 
	\cos\left( \frac{\sum^{M}_{m=0}(R^{-1})^m_{299}r_m a^{299}}{f_a} +\Theta\right) -\frac{\kappa}{n^2}
	\cos\left( \frac{n\sum^{M}_{m=0}(R^{-1})^m_{299}r_m a^{299}}{f_a} \right)\right] \,.
\end{align}
Using the initial and final field values reported in
\Cref{tab:observablescomparison}, we find that this effective
single-field description reproduces the same cosmological observables
as the full multifield dynamics. The only noticeable difference arises
in the total number of e-folds, which is slightly reduced,
$N_\star \simeq 38.5$, compared to the multifield case.
This agreement justifies the use of the single-field approximation for
describing the inflationary dynamics. Moreover, it provides a convenient
framework to disentangle the inflationary evolution from the subsequent
cosmological epochs, in particular when analyzing the \ac{DM}
dynamics discussed in the following sections.

\subsection{Reheating}
\label{sec:reheating}

During inflation, the dynamics are dominated by the heaviest field,
$a^{299}$, while the remaining fields remain effectively frozen, as
illustrated in \Cref{fig:dynamicseigenstates}. After inflation, the
inflaton field $a^{299}$ begins to oscillate around the minimum of its
potential, and reheating proceeds through its coupling to photons, as
given in \cref{eq:4-d_action}. As a result, the Universe transitions
from an initial phase dominated by the coherent oscillations of the
inflaton to the standard radiation-dominated era, at a time
$t_{\mathrm{RH}}$ corresponding to a reheating temperature
$T_{\mathrm{RH}}$.
The structure of the leading contributions to the potential plays a
crucial role in determining the decay dynamics. In contrast, the lighter
fields remain frozen, with their energy densities behaving effectively as
vacuum energy, until the time $t_{i,\mathrm{osc}}$ at which the condition
$3H(t_{i,\mathrm{osc}}) \simeq 2 m_{i}$ is satisfied. After this point, 
they begin to oscillate.

At the end of inflation, $a^{299}$ rolls towards its minimum, whereas the other
fields remain close to their initial values. This motivates expanding
the potential~\eqref{eq:V299} around the minimum for $a^{299}$, keeping the leading terms
\begin{equation}
V(a^{299}) ~\simeq~
\frac{1}{2} m_{299}^2 \left(a^{299} - a^{299}_{\mathrm{min}}\right)^2 
+ \lambda_{299} \left(a^{299} - a^{299}_{\mathrm{min}}\right)^4 \,,
\end{equation}
where the parameters are given in \Cref{tab:bestfit}. At early times,
the quartic term dominates the dynamics, while at later times the quadratic
term becomes dominant. As discussed above, we define the end of inflation by the condition
$\ddot{\mathrm{a}}=0$, which is equivalent to $w=-1/3$, or
$(\dot{a}_{\mathrm{end}}^{299})^2 = V\left(a^{299}_\mathrm{end}\right)$.
The corresponding energy density of the inflaton at the end of inflation is
\begin{equation}\label{eq:Vend}
\rho_\mathrm{end} ~=~ \frac{3}{2} V\left(a^{299}_\mathrm{end}\right) \, ,
\end{equation}
where the total energy density is dominated by the inflaton field $a^{299}$
during inflation, as can be explicitly verified within this model.
The post-inflationary evolution of the energy density of $a^{299}$ is
described by
\begin{equation}\label{eq:rho_299eom}
\frac{\dd \rho_{299}}{\dd t} + 3(1+w) H \rho_{299} ~=~ - \Gamma_{299} (1+w)\, \rho_{299} \,,
\end{equation}
where $\Gamma_{299}$ denotes its decay rate. The equation-of-state parameter
$w$ depends on the shape of the potential and is given by
$w = \frac{k-2}{k+2}$ for a potential of the form $V(\phi)\propto \phi^k$. As the potential
is first dominated by $k=4$ the field $a^{299}$ behaves as radiation
and then when the field $k=2$ the field $a^{299}$ decays as matter.
During the oscillatory phase, the field evolution can be approximated as~\cite{Garcia:2020wiy,Clery:2024dlk}
\begin{equation}
    a^{299}(t)~\simeq~ \tilde{a}^{299}(t)\,\mathcal{P}(t) \, ,
\end{equation}
where $\tilde{a}^{299}(t)$ denotes the slowly varying envelope of the
oscillations, determined from the averaged energy density as
$\langle \rho_{299} \rangle \simeq \langle V(a^{299}) \rangle \simeq V(\tilde{a}^{299})$,
while $\mathcal{P}(t)$ is a quasiperiodic function encoding the oscillatory
behavior in the potential. As the amplitude $\tilde{a}^{299}$ decreases, the quadratic contribution
to the potential eventually becomes comparable to the quartic term at $\mathrm{a}=\mathrm{a}_m$ and
subsequently dominates the dynamics. The transition occurs when
\begin{equation}
    \frac{1}{2} m_{299}^2
    \left(\tilde{a}^{299}(\mathrm{a}_m)-a^{299}_{\mathrm{min}}\right)^2
    ~=~
    \lambda_{299}
    \left(\tilde{a}^{299}(\mathrm{a}_m)-a^{299}_{\mathrm{min}}\right)^4 \, .
\end{equation}
During the quartic-dominated phase, the energy density scales as $\rho_{299}\simeq \rho_{ \mathrm{end}}\left(\frac{\mathrm{a}_\mathrm{end}}{\mathrm{a}}\right)^4$, which implies that the oscillation 
envelope evolves as $\tilde{a}^{299} \simeq a^{299}_\mathrm{end}\left(\frac{\mathrm{a}_\mathrm{end}}{\mathrm{a}}\right)$.
Substituting this relation into the previous condition gives
\begin{equation}
\mathrm{a}_m ~=~ \sqrt{2\lambda }\frac{|a^{299}_\mathrm{end}-a^{299}_{\mathrm{min}}|}{m_{299}} \mathrm{a}_\mathrm{end}\,,
\end{equation}
which, as we shall see in \Cref{tab:scalefactor}, is small enough in our scenario to avoid fragmentation considerations~\cite{Clery:2024dlk}.
Having determined the transition point, we can now solve
\cref{eq:rho_299eom} in two regimes. First, during the
quartic-dominated phase, for $\mathrm{a}_{\mathrm{end}} \leq \mathrm{a} \leq \mathrm{a}_m$,
where the equation of state parameter is $w=1/3$. Subsequently, for
$\mathrm{a}_m < \mathrm{a} < \mathrm{a}_{\mathrm{RH}}$, the quadratic term dominates with
a corresponding equation of state $w=0$.
Here, $\mathrm{a}_{\mathrm{RH}}$ denotes the scale factor at the end of
reheating, which we define through the condition
$\rho_{299}(\mathrm{a}_{\mathrm{RH}}) =\rho_r(\mathrm{a}_{\mathrm{RH}})$.
The energy density equation~\eqref{eq:rho_299eom} in terms of 
the scale factor is given by
\begin{equation}
\frac{\dd \rho_{299}}{\dd \mathrm{a}} + \frac{3(1+w)}{\mathrm{a}}  \rho_{299}
~=~ - \frac{\Gamma_{299} (1+w)\, \sqrt{3\rho_{299}}\, M_\mathrm{Pl}}{\mathrm{a}} \,,
\end{equation}
where we have assumed that $H^2\simeq \rho_{299}/3 M_\mathrm{Pl}^2$ as expected during reheating epoch. 
This equation has the analytical solution
\begin{equation}\label{eq:rho299sol} 
\rho_{299}~=~\frac{1}{3} \mathrm{a}^{-3 (w+1)} \left(M_\mathrm{Pl}^2\Gamma_{299}^2  \mathrm{a}^{3 w+3}-2 M_\mathrm{Pl}\,\Gamma_{299} \mathrm{a}^{\frac{3 (w+1)}{2}} \sqrt{\mathrm{a}_\mathrm{init}^{3 w+3} \Delta}+\mathrm{a}_\mathrm{init}^{3 w+3}\Delta \right) \, ,
\end{equation}
where we defined
\begin{equation}
    \Delta ~:=~ M_\mathrm{Pl}^2\Gamma_{299}^2 -2 \sqrt{3\rho_{299, \mathrm{init}}}\, M_\mathrm{Pl}\Gamma_{299} +3 \rho_{299, \mathrm{init}}\, .
\end{equation}
Here, $\mathrm{a}_{\mathrm{init}}$ denotes the initial scale factor for each
dynamical regime, namely $\mathrm{a}_{\mathrm{init}}=\mathrm{a}_{\mathrm{end}}$ during the
quartic-dominated epoch and $\mathrm{a}_{\mathrm{init}}=\mathrm{a}_m$ during the
quadratic-dominated epoch, while $\rho_{299, \mathrm{init}}:=\rho_{299}(\mathrm{a}_{\mathrm{init}})$. The equation-of-state parameter takes the
corresponding values $w=1/3$ and $w=0$ in each case.
For calculating
the decay rate we are going to assume that the quadratic term dominates for most of 
its evolution. Considering the coupling to photons given by \cref{eq:4-d_action},
in the mass eigenbasis we have
\begin{equation}
     \mathcal{L}_{a^{299}, \mathrm{int}}~=~ -\frac{1}{4}g_{a\gamma\gamma}\sum_{m=0}^{M} (R^{-1})^m_{299}r_m a^{299}F_{\mu\nu}\widetilde{F}^{\mu\nu} \, .
\end{equation}
We can then define an effective coupling constant for $a^{299}$ which is given 
by $\tilde{g}_{a\gamma\gamma}= g_{a\gamma\gamma}\sum_{m=0}^{M} (R^{-1})^m_{299}r_m \simeq 24.5g_{a\gamma\gamma}$ with
$g_{a\gamma\gamma}$ shown in \Cref{tab:bestfit}. With this effective coupling constant, we can calculate
the decay rate from $a^{299}$ to a pair of photons given by~\cite{Kim:2008hd}
\begin{equation}
\Gamma (a^{299}\rightarrow \gamma \gamma) ~=~ \frac{\tilde{g}_{a\gamma\gamma}^2 m_{299}^3}{64\pi}
~=~ \left(\sum_{m=0}^{M} (R^{-1})^m_{299}r_m\right)^2\frac{\alpha^2 c_\gamma^2m_{299}^3}{64\pi^3 f_a^2} \, ,
\end{equation}
where we have defined
\begin{equation}
g_{a\gamma\gamma} ~:=~ \frac{c_\gamma \alpha}{\pi f_a} \, ,
\end{equation}
with $\alpha$ denoting the fine-structure constant and $c_\gamma$ a model-dependent dimensionless coefficient.
Given this decay rate, we can also determine the evolution of the radiation
energy density. The radiation energy density
has source terms involving every field of our ensemble. So
the corresponding continuity equation reads
\begin{equation}\label{eq:radeq}
	\frac{\dd \rho_{r}}{\dd t} + 4 H \rho_r
	~=~ \sum_{i=0}^{M}\Gamma_{i} (1+w_i)\, \rho_{i} \, ,
\end{equation}
where $w_i$ is the equation of state for each field $a^i$. 
At early times before reheating time $t\ll t_\mathrm{RH}$, the source term of the inflaton is expected 
to dominate, that is
\begin{equation}\label{eq:rehdecaycondition}
\Gamma_{299}\rho_{299} ~\gg~ \sum_{i=0}^{298}\Gamma_i \rho_i \, .
\end{equation}
\begin{figure}
    \centering
    \includegraphics[width=0.65\linewidth]{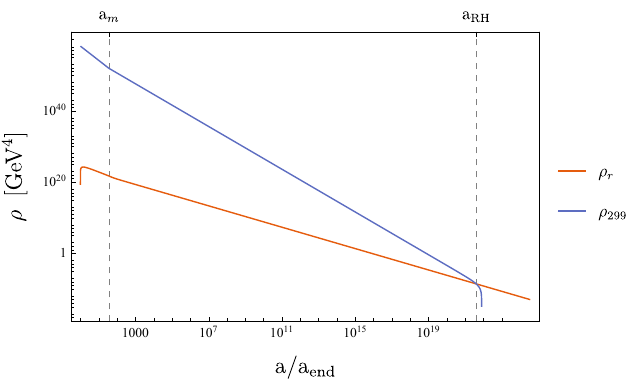}
    \caption{ Evolution of the inflaton energy density $\rho_{299}$ and the radiation energy density $\rho_r$  during the reheating epoch. The vertical lines represent the transition from the quartic to quadratic epoch $\mathrm{a}_m$ and the time at
    which reheating ends $\mathrm{a_{RH}}$. }
    \label{fig:RH299andrad}
\end{figure}
Under these assumptions, \cref{eq:radeq} can be rewritten in terms
of the scale factor as
\begin{equation}\label{eq:rhorscalefactor}
\frac{\dd \rho_{r}}{\dd \mathrm{a}} + \frac{4}{\mathrm{a}}\, \rho_{r}
= \frac{\Gamma_{299} (1+w)\, \sqrt{3\rho_{299}}\, M_\mathrm{Pl}}{\mathrm{a}} \, ,
\end{equation}
where we have assumed that the total energy
density is dominated by $\rho_{299}$. Using $\rho_{299}\simeq \rho_{299, \mathrm{init}}(\mathrm{a_{init}}/\mathrm{a})^{3+3w}$, we obtain the solution
to \cref{eq:rhorscalefactor} as
\begin{equation}
    \small
    \rho_r=\frac{\mathrm{a}_{\text{init}}^4 \left(2 \sqrt{3} M_\mathrm{Pl}\Gamma_{299} (w+1) \sqrt{\rho_{299, \text{init}}}+(3 w-5) \rho_{\text{r,init}}\right)-2 \sqrt{3} \mathrm{a}^4 M_\mathrm{Pl}\Gamma _{299} (w+1) \sqrt{\mathrm{a}^{-3 (w+1)} \rho_{299, \text{init}} \mathrm{a}_{\text{init}}^{3 w+3}}}{\mathrm{a}^4 (3 w-5)} \, ,
\end{equation}
with $\rho_{r, \mathrm{init}}=0$ for $\mathrm{a_{end}}\leq \mathrm{a} \leq \mathrm{a}_m$ and $\rho_{r, \mathrm{init}}=\rho_{r}(\mathrm{a_m})$ for $\mathrm{a}>\mathrm{a}_m$. This allows us to
determine the reheating scale factor $\mathrm{a}_{\mathrm{RH}}$ from the condition
$\rho_{\mathrm{RH}} ~:=~ \rho_{299}(\mathrm{a}_{\mathrm{RH}})
= \rho_r(\mathrm{a}_{\mathrm{RH}})$.
The corresponding reheating temperature is then given by
\begin{equation}
    T_{\mathrm{RH}} =
    \left(\frac{30\,\rho_{\mathrm{RH}}}{\pi^2 g_{\mathrm{RH}}}\right)^{1/4} \, .
\end{equation}
For the parameters of our model, we find
$\mathrm{a}_{\mathrm{RH}}/\mathrm{a}_\mathrm{end} = 3.97 \times 10^{21}$ and
$\rho_{\mathrm{RH}} = 2.61 \times 10^{-9}\,\mathrm{GeV}^4$,
which corresponds to a reheating temperature of
$T_{\mathrm{RH}} = 5.21\,\mathrm{MeV}$. This value lies close to the
lower bound required for consistency with \ac{BBN}, typically
$T_{\mathrm{RH}} \gtrsim \mathcal{O}(1)\,\mathrm{MeV}$~\cite{Fields:2019pfx, Hasegawa:2019jsa}.

So far, we have neglected the contributions of the remaining fields $a^i$ with $i<M$,
assuming that their dynamics are not relevant during the reheating epoch. As argued above, and 
as we will show explicitly below, this is justified because the energy densities of these 
additional fields never dominate the total energy density during this stage, and 
the condition in \cref{eq:rehdecaycondition} remains satisfied.
However, the onset of oscillations of these fields determines the initial conditions 
relevant for the dark matter abundance at late times. 
This aspect will be discussed in detail in \Cref{sec:DM}.

\section{Spectral distortions}
\label{sec:spectral}

As discussed above, our model can be approximated as an effective single-field scenario. 
Within this framework, we compute the \acp{SD} generated in the
\ac{CMB}, accounting for the dissipation of primordial
acoustic waves with adiabatic initial conditions~\cite{Hu:1994bz,Chluba:2019nxa}.

The photon intensity spectrum at redshift $z$ can be expressed as the sum of a 
blackbody distribution and a small distortion component,
\begin{equation}
\label{eq:intensity_decomposition}
   I(z,x) ~=~ I_0(x)+\Delta I(z,x)\,,
\end{equation}
where $x:= h\nu/(k_B T)$ is the dimensionless frequency and $I_0(x)$ denotes the Planck spectrum. Since the distortions are small ($\Delta I/I\ll1$), the Boltzmann equation for the photon phase-space distribution can be linearized around the blackbody solution. The resulting distortion can then be written using a Green's function approach~\cite{Lucca:2019rxf},
\begin{equation}
\label{eq:green_solution}
  \Delta I(z,x) ~=~ \Delta I_R(z,x) +\int_z^{\infty} \frac{\dd Q(z')/\dd z'}{\rho_\gamma(z')}\,G_{\rm th}^\star(x,z')\,\dd z',
\end{equation}
where $\rho_\gamma^{-1} \dd Q/\dd z$ is the effective heating rate,
$G_{\rm th}^\star$ is the thermalization kernel encoding the relevant microphysics, 
and $\Delta I_R$ denotes the residual term arising from the linearization procedure. 

Spectral distortions originate from the damping of small-scale acoustic modes after 
horizon re-entry. Pressure gradients drive acoustic oscillations, while photon 
diffusion (Silk damping) dissipates their energy, injecting heat into the photon bath 
and generating \acp{SD}~\cite{Chluba:2013dna}. The corresponding effective heating rate can be approximated as
\begin{equation}
\label{eq:heating_rate}
  \frac{1}{\rho_\gamma}\frac{\dd Q}{\dd z} ~=~ 4A^2 \int_{k_{\min}}^\infty \frac{k^4\,\dd k}{2\pi^2}\,P_{\mathcal R}(k)\,\partial_z k_D^{-2}\,\e^{-2k^2/k_D^2}\,,
\end{equation}
where $P_{\mathcal R}(k)$ is the primordial curvature power spectrum derived from the 
inflationary potential $V(a^{299})$ defined in \cref{eq:V299}. For adiabatic perturbations,
\begin{equation}
  A ~\simeq~ \left(1+\frac{4R_\nu}{15}\right)^{-1}\,,
\end{equation}
which accounts for the effect of neutrino anisotropic stress. The photon diffusion scale is 
\begin{equation}
  k_D ~\simeq~ \num{4.048e-6} (1+z)^{3/2}\,\mathrm{Mpc}^{-1},
\end{equation}
and we set $k_{\min}=1\ \mathrm{Mpc}^{-1}$.

Since the heating rate depends directly on $P_{\mathcal R}(k)$, spectral distortions 
provide a powerful probe of primordial fluctuations and, consequently, of inflationary 
physics~\cite{Henriquez-Ortiz:2022ulz,Baur:2023naq}. To avoid potential systematic uncertainties associated with the residual term $\Delta I_R$, particularly for larger distortions, we compute the full \ac{SD} signal numerically using the \texttt{CLASS} code.

\begin{figure}[t]
	\centering
	\includegraphics[width=0.45\textwidth]{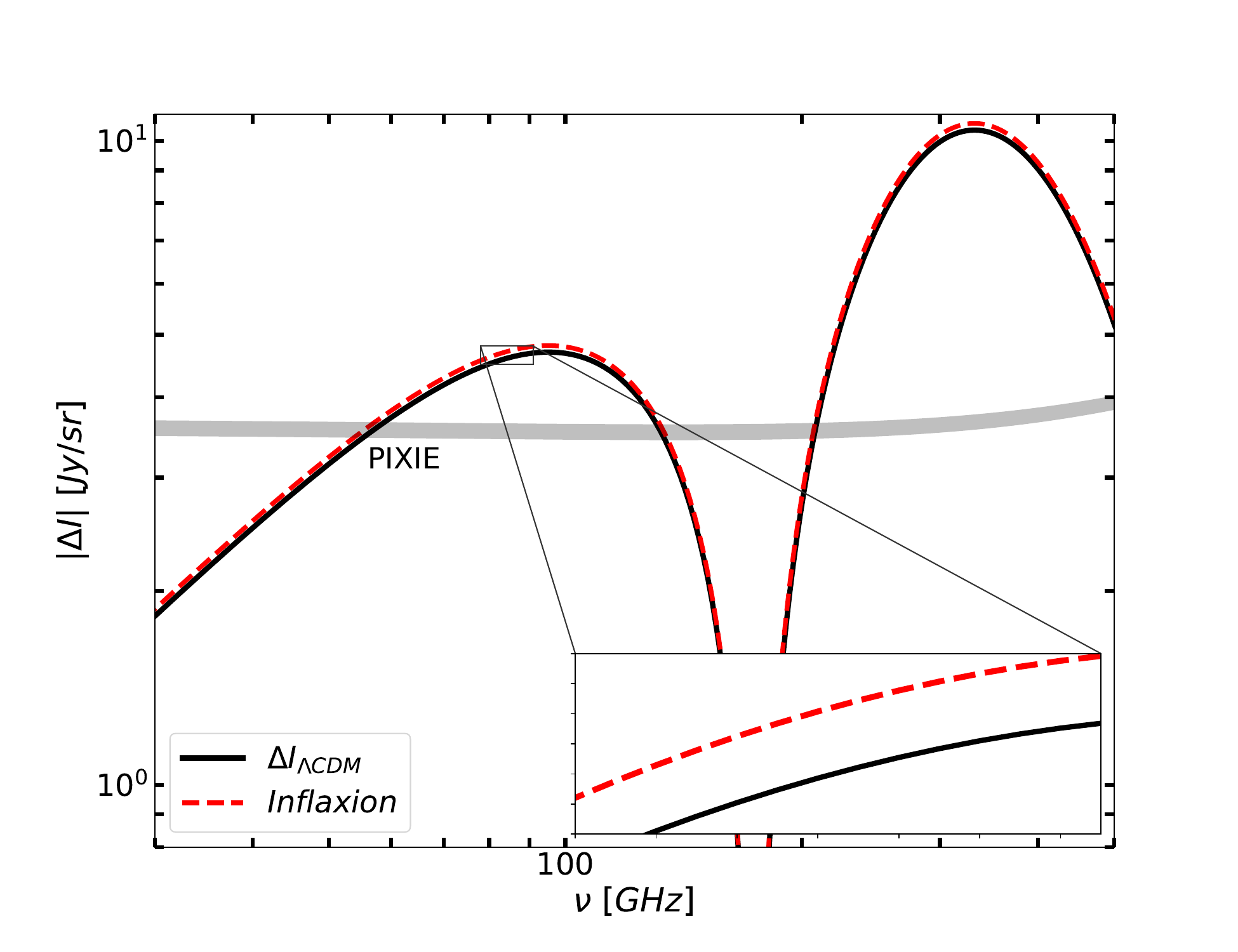}
	\includegraphics[width=0.45\textwidth]{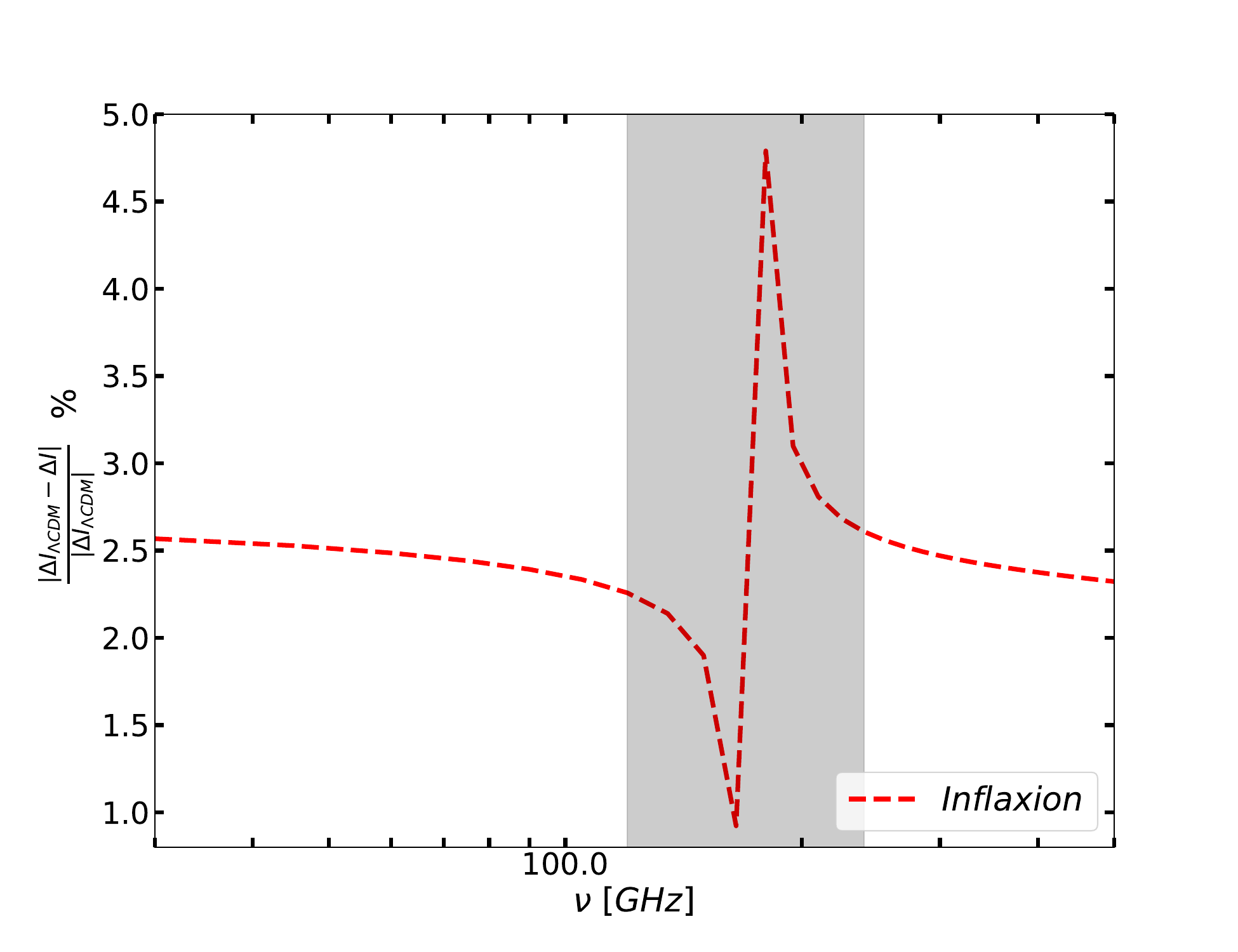}
	\caption{\label{fig:SD}
	Left: total spectral distortion $\Delta I$ predicted by our model (red curve), 
	compared with the $\Lambda$CDM prediction (dark gray region) and the expected 
	sensitivity of a future PIXIE-like experiment (light gray shaded region). 
	Right: percentage difference between the signal predicted by our model and the 
	$\Lambda$CDM prediction. For frequencies outside the range 
	$100~\mathrm{GHz} \lesssim \nu \lesssim 250~\mathrm{GHz}$, the difference with 
	respect to the fiducial signal is approximately $2.5\%$.}
\end{figure}

The left panel of \Cref{fig:SD} shows the spectral distortion signal predicted in the
single-field limit of our model (red curve). Relative to the $\Lambda$CDM prediction, 
the \ac{SD} signal is suppressed by approximately $2.5\%$ at frequencies 
$\nu \lesssim 100~\mathrm{GHz}$ and $\nu \gtrsim 250~\mathrm{GHz}$, as highlighted in the 
right panel.

This suppression can be directly traced back to small differences in the primordial 
curvature power spectrum $P_{\mathcal R}(k)$ at scales relevant for diffusion damping. 
Since the heating rate \cref{eq:heating_rate} is weighted toward modes with
$k \sim k_D(z)$, even mild changes in the amplitude or scale dependence of 
$P_{\mathcal R}(k)$ at $k \gtrsim 1\,\mathrm{Mpc}^{-1}$ lead to a cumulative reduction 
in the injected energy and, consequently, in the \ac{SD} signal.

Physically, this reflects a slightly less efficient dissipation of small-scale acoustic 
modes in our model compared to $\Lambda$CDM. Although the effect is modest, it is 
systematic across frequencies and therefore constitutes a potentially observable 
signature. In particular, the percent-level deviation lies within the sensitivity range 
of future PIXIE-like experiments, suggesting that spectral distortions could provide a 
complementary probe of the small-scale primordial spectrum and help discriminate 
between inflationary scenarios that are otherwise indistinguishable at CMB anisotropy 
scales.

\section{Dark Matter}
\label{sec:DM}

As discussed previously, the field $a^M$ can be identified with the
inflaton, while the remaining fields $a^i$ with $i<M=299$ behave effectively
as vacuum energy until they begin to oscillate after the end of
inflation. The onset of oscillations for each field occurs at a time
$t_{i,\mathrm{osc}}$ determined by the condition
$3H(t_{i,\mathrm{osc}})\simeq 2m_{i}$. Prior to this transition, the
corresponding fields do not contribute to the total \ac{DM}
abundance $\Omega_{\mathrm{CDM}}$, but instead behave as an effective
dark energy component. Consequently, the initial field displacements
play a crucial role in determining the energy densities that govern the
subsequent cosmological evolution.

\subsection{Initial conditions}

We assume that the axion population is generated through non-thermal
processes, in particular via the misalignment mechanism, as in the
\ac{DDM} scenario~\cite{Dienes:2011sa}. In this framework, the
instanton-induced potential in \cref{eq:Vinstanton} is assumed to
be absent at temperatures $T \gg \Lambda$, where non-perturbative
effects are negligible. Prior to the onset of these effects, the
five-dimensional field $\Phi$ is effectively massless and may therefore
acquire an arbitrary initial \ac{VEV}.
Once the non-perturbative potential is generated, the relevant quantity is the displacement of the field from the minimum of the resulting potential given by \cref{eq:minphi0,eq:minphim}. We therefore parametrize the initial field displacement in terms of the misalignment angle $\theta$ as
\begin{equation}
\langle \Phi \rangle -\vev{\Phi_\mathrm{min}} ~=~ \theta f^{3/2}_\Phi\,,
\end{equation}
with $\theta \in [-\pi,\pi]$. After dimensional reduction, as described
in \cref{eq:expansion}, the resulting four-dimensional fields $\phi^n$
with $n\neq 0$ acquire potentials determined by their corresponding
\ac{KK} masses, whose minima are located at $\phi^n=0$. In contrast,
the zero mode $\phi^0$ remains massless and receives no contribution to
its potential at this stage. Consequently, there is no preferred
minimum selecting its initial value, and the corresponding \ac{VEV} can
be treated as arbitrary. The induced initial \ac{VEV} 
is then given by
\begin{equation}\label{eq:vevphi0}
\vev{\phi^0}-\vev{\phi^0_\mathrm{min}} ~=~ \sqrt{2\pi R}\,\theta f^{3/2}_\Phi\,,\quad\qquad
\vev{\phi^n} ~=~ 0 \;\;\text{for} \;\; n\neq 0 .
\end{equation}
After the change of basis to the mass eigenbasis, the expected 
values of the mass eigenstates are
\begin{align}
\vev{a^m}-\vev{a^m_\mathrm{min}} ~=~ \sum_{n=0}^{M} R^{m}_{n}\,\left( \vev{\phi^n} - \vev{\phi^n_\mathrm{min}} \right)\, ,
\qquad m=0,\ldots,M \, ,
\end{align}
where $R^{m}_{n}$ is the rotation matrix element.
The expectation values in \cref{eq:vevphi0} imply that
\begin{align}\label{eq:vevrotation}
\vev{a^m}-\vev{a^m_\mathrm{min}} ~=~ R_0^m\, (\vev{\phi^0}-\vev{\phi^0_\mathrm{min}})\,  , \qquad  m=0,\ldots,M \, . 
\end{align}
Since the fields remain frozen at their initial values before the
appearance of non-perturbative effects, and the potential for every
field $a^i$ other than the inflaton is dominated by the quadratic
contribution, the corresponding initial energy density stored in each
field is given by
\begin{equation}
\label{eq:initenergydensities}
   \rho_{i,\mathrm{osc}} := \rho_i(t_{i,\mathrm{osc}}) =    \frac{1}{2}m_{i}^2 \left(\vev{a^i}-\vev{a^i_\mathrm{min}} \right)^2
    =     \frac{1}{2}m_{i}^2 (R_0^i)^2 \left(\vev{\phi^0}-\vev{\phi^0_\mathrm{min}} \right)^2
    =     \frac{1}{2}m_{i}^2 (R_0^i)^2 f_a^2 \theta ^2
    \,,
\end{equation}
recalling that $f_a= \sqrt{2\pi R} f^{3/2}_\Phi$.
This corresponds to the energy density at the time the fields begin to
oscillate, before subsequently redshifting as non-relativistic matter.

It is then important to determine during which cosmological epoch these
fields enter their oscillatory regime. This can be inferred from the Hubble parameter at the beginning of the
radiation-dominated era, corresponding to the reheating temperature
$T_{\mathrm{RH}}$. For our model, we obtain
$H_{\mathrm{RH}} = \num{1.21e-23}\,\mathrm{GeV}$,
which is much smaller than the mass of the lightest field,
$m_{0} = \num{7.86e6}\,\mathrm{GeV}$.
This hierarchy implies that all lighter fields satisfy the condition
$3H(t_{i,\mathrm{osc}}) \simeq 2 m_{i}$ well before the radiation era
begins, and therefore start oscillating already during the reheating
phase. Consequently, the majority of the ensemble behaves as
pressureless matter prior to the completion of reheating.
Nevertheless, as we will show below, this does not modify the analysis
presented in \Cref{sec:Inflation}, since the energy density of these
fields never dominates the total energy during this epoch.

\subsection{DM during reheating}

\begin{figure}
	\centering
	\includegraphics[width=0.75\textwidth]{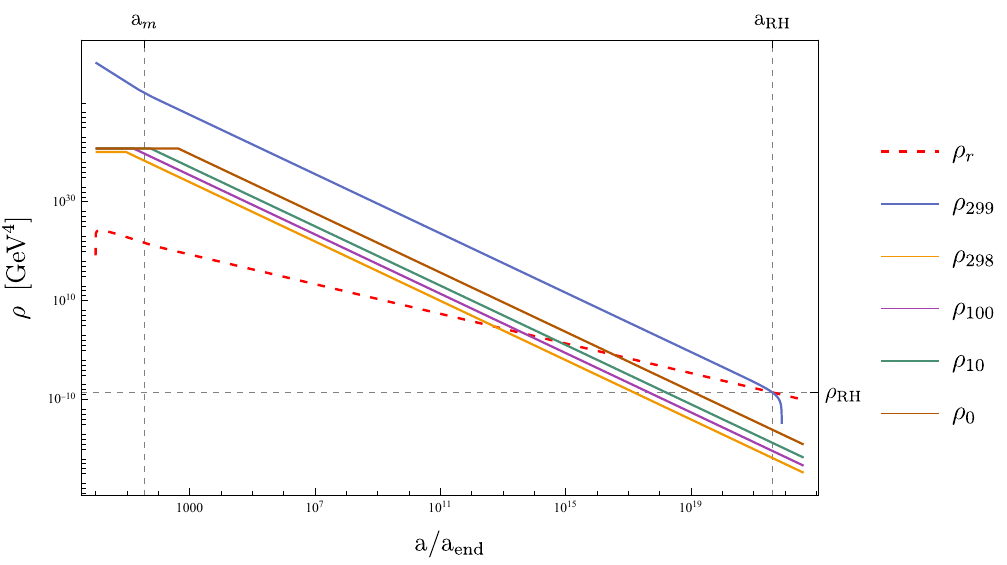}
	\caption{\label{fig:reheating}
Evolution of the energy densities $\rho_i$ associated with the fields
$a^i$ during the reheating epoch, together with the radiation energy
density $\rho_r$, indicated by a red dashed line. 
The figure shows that the energy densities of the
\ac{DM} fields remain subdominant throughout reheating.
	}
\end{figure}

Once the condition $3H(t_{i,\mathrm{osc}})\simeq 2m_{i}$ is satisfied,
the corresponding field $a^i$ begins to oscillate around the minimum of
its quadratic potential. In this regime, the field behaves as
pressureless matter with equation-of-state parameter $w=0$. During this
stage, the energy density of each field evolves according to the
continuity equation
\begin{equation}
\label{eq:DMcontinuityeqs}
	\frac{\dd \rho_i}{\dd t} + 3 H \rho_i ~=~ -\Gamma_i\rho_i \, , 
\end{equation}
where we have used $w=0$ for all \ac{DM} fields. Throughout the
reheating epoch, the Hubble parameter is approximately determined by
the inflaton energy density,
$H^2\simeq \rho_{299}/3M_\mathrm{Pl}^2$.
The initial conditions for the energy densities are given in
\cref{eq:initenergydensities}.
Writing \cref{eq:DMcontinuityeqs} in terms of the scale
factor takes the form
\begin{equation}
\label{eq:scalefactorDM}
  \frac{\dd \rho_i}{\dd \mathrm{a}}  +\frac{3}{\mathrm{a}}\rho_i ~=~ -\frac{\Gamma_i\rho_i}{\mathrm{a}H} \, .
\end{equation}
If one assumes that the Hubble parameter is always dominated by one of the components of the Universe and
neglected the contribution from the decays, that is
\begin{equation}
H\simeq \sqrt{\frac{\rho_{H, \mathrm{init}}\left(\frac{\mathrm{a}_{H,\mathrm{init}}}{\mathrm{a}} \right)^{3(1+w)}}{3 M^2_\mathrm{Pl}}}   
\end{equation}
we find the general solution of \cref{eq:scalefactorDM}, given by
\begin{equation}\label{eq:rhoisol}
\rho_i(\mathrm{a})=
\rho_{i,\mathrm{init}}
\left(\frac{\mathrm{a}_{\mathrm{init}}}{\mathrm{a}}\right)^3
\exp\!\left[
-\frac{2M_\mathrm{Pl}\Gamma_i}
{\sqrt{3\rho_{H,\mathrm{init}}}\,(w+1)}
\left(
\frac{\mathrm{a}_{\mathrm{init}}}
{\mathrm{a}_{H,\mathrm{init}}}
\right)^{\frac{3}{2}(w+1)}
\left(
\left(
\frac{\mathrm{a}}
{\mathrm{a}_{\mathrm{init}}}
\right)^{\frac{3}{2}(w+1)}
-1
\right)
\right].
\end{equation}
It is important to distinguish $\mathrm{a}_{H,\mathrm{init}}$ from
$\mathrm{a}_{\mathrm{init}}$, as these quantities generally refer to
different physical events. The quantity
$\mathrm{a}_{H,\mathrm{init}}$ denotes the scale factor at which the
energy density component dominating the Hubble expansion is parameterized.
In our case, this corresponds either to
$\mathrm{a}_{\mathrm{end}}$, when the inflaton evolution is governed by
the quartic part of its potential, or to $\mathrm{a}_m$, marking the
transition to the quadratic-dominated regime. By contrast,
$\mathrm{a}_{\mathrm{init}}$ specifies the initial condition for the
evolution of a given field $a^i$. Depending on the epoch under
consideration, it is given either by the onset of oscillations,
$\mathrm{a}_{i,\mathrm{osc}}$, or by the transition point
$\mathrm{a}_m$. The two quantities coincide only when the field begins
oscillating precisely at the same scale factor used to normalize the
dominant contribution to the Hubble parameter.
In obtaining this solution, it is therefore necessary to distinguish
between fields that begin oscillating before the transition point
$\mathrm{a}_m$, when the inflaton potential is dominated by its quartic
term, and those that start oscillating during the subsequent
quadratic-dominated phase.
The full evolution is therefore obtained by matching the
solutions in \cref{eq:rhoisol} through the appropriate boundary conditions at $\mathrm{a}=\mathrm{a}_m$. The solution for those fields that oscillate
before the quadratic transition $\mathrm{a}_m$ is given by
\begin{equation}
 \rho_i ~=~  \rho_{i,\mathrm{osc}} \left(\frac{\mathrm{a}_{i, \mathrm{osc}}}{\mathrm{a}}\right)^3 \cdot \begin{cases}
 \exp\!\left[-3\,\beta_i\,\mathrm{a}_{i,\rm osc}^2\left(\frac{\mathrm{a}^2}{\mathrm{a}_{i,\mathrm{osc}}^2}-1\right)
\right]\, ,  & \mathrm{a_{i, osc}}\leq \mathrm{a} \leq \mathrm{a}_m\, , \\[1em]

    \exp\!\left[-\beta_i\,\mathrm{a}_{m}^2 \left(4\left(\frac{\mathrm{a}}{\mathrm{a}_m}\right)^{3/2}-1-3\left(\frac{\mathrm{a}_{i,\mathrm{osc}}}{\mathrm{a}_m}\right)^2\right)\right]\, , & \mathrm{a}_m \leq \mathrm{a} \leq \mathrm{a_{RH}}\, , 
 \end{cases}  
\end{equation}
while for those fields that oscillate after the quadratic transition
is given by
\begin{equation}
 \rho_i ~=~ \rho_{i,\mathrm{osc}} \left(\frac{\mathrm{a}_{i, \mathrm{osc}}}{\mathrm{a}}\right)^3 \exp\!\left[-4\beta_i\,\sqrt{\mathrm{a}_m\mathrm{a}_{i, \mathrm{osc}}^{3}}\left(\left(\frac{\mathrm{a}}{\mathrm{a_{i,\mathrm{osc}}}}\right)^{3/2}-1\right)\right]\, , \quad\quad \mathrm{a_{i, osc}}\leq \mathrm{a} \leq \mathrm{a}_\mathrm{RH} \, , 
\end{equation}
where we have defined 
\begin{equation}
	\beta_i~=~ \frac{M_\mathrm{Pl} \Gamma_i}{2\sqrt{3\rho_{\mathrm{end}}}\mathrm{a}_{\mathrm{end}}^2}\, .
\end{equation}
The resulting evolution of the energy densities as functions of the
scale factor is shown in \Cref{fig:reheating}, together with the
previously determined inflaton and radiation energy densities. For the
\ac{DM} fields, we adopt the initial conditions discussed in the next
subsection, chosen such that the total relic abundance of the ensemble
reproduces the observed \ac{DM} abundance today. As can be seen,
the \ac{DM} components remain subdominant throughout the reheating epoch
and never dominate the total energy density of the Universe. This
behavior validates the assumptions made in the previous reheating
analysis, where the contribution of the \ac{DM} sector to the
background evolution was neglected.

\begin{figure}
	\centering
	\includegraphics[width=0.65\textwidth]{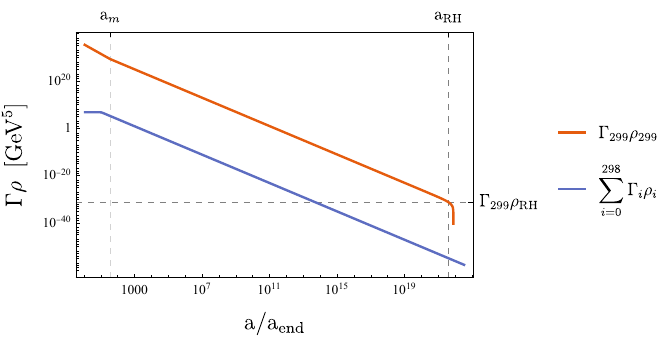}
	\caption{\label{fig:decaysdomination}
Comparison between the decay source term of the inflaton,
$\Gamma_{299}\rho_{299}$, and the combined contribution from the
\ac{DM} sector, $\sum_{i=0}^{M-1} \Gamma_i \rho_i$. The figure shows that
the inflaton contribution dominates throughout the reheating epoch,
implying that the evolution of the radiation energy density is
primarily sourced by the decay of the inflaton field.
	}
\end{figure}

Using these solutions, we can also evaluate the decay source terms
appearing in the radiation continuity equation,
\cref{eq:radeq}. In \cref{sec:Inflation}, we assumed that the
dominant contribution to the radiation source term arises from the
inflaton decay. This assumption can be explicitly verified in
\Cref{fig:decaysdomination}, where we show that the combined
contribution from the \ac{DM} fields never exceeds the decay source term
associated with the inflaton field $a^{299}$. Consequently, the
approximation used in deriving the evolution of the radiation energy
density remains valid throughout the reheating epoch.

\subsection{DM abundances}

\begin{figure}
    \centering
    \includegraphics[width=0.65\linewidth]{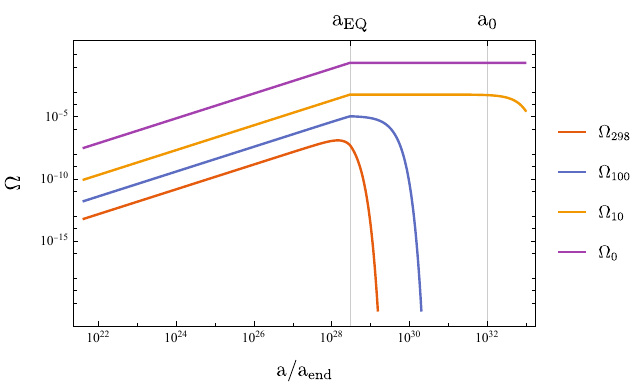}
    \caption{\label{fig:abundances}
    Evolution of the individual abundances $\Omega_i$ associated with the
\ac{DM} fields during the radiation and matter dominated epochs. The
figure also indicates the scale factor at matter-radiation equality,
$\mathrm{a}_{\mathrm{EQ}}$, together with the present-day scale factor. The late-time
behavior shows how the individual components contribute to the total \ac{DM} abundance today. }
\end{figure}

Having established the initial conditions at the end of reheating, we
can now study the cosmological evolution of the \ac{DM} abundances
during the radiation- and matter-dominated eras. In particular, we are
interested in determining the present-day total \ac{DM} abundance
$\Omega_{\mathrm{DM},0}$ arising from the full ensemble.
The abundance associated with each field $a^i$ is defined as
\begin{equation}
    \Omega_i
    ~=~
    \frac{\rho_i}{3H^2 M_\mathrm{Pl}^2} \, ,
\end{equation}
where $\rho_i$ is obtained by solving
\cref{eq:DMcontinuityeqs} with initial conditions determined by the
energy densities at the end of reheating. The total \ac{DM}
abundance is then given by
\begin{equation}
    \Omega_{\mathrm{DM}}
    ~=~
    \sum_i \Omega_i \, .
\end{equation}
In solving for the subsequent evolution, we approximate the Hubble
parameter using the dominant component of the energy density in each
cosmological epoch. 
Assuming that the expansion is successively dominated by the quartic inflaton potential,
the quadratic inflaton potential, radiation, and finally the total matter energy density
(including \ac{DM} and baryons), the Hubble parameter evolves according to
\begin{equation}
	H^2 ~\simeq~ \frac{\rho_{ \mathrm{end}}}{3M^2_\mathrm{Pl}} \cdot \begin{cases}  \left(\frac{\mathrm{a}_{\mathrm{end}}}{\mathrm{a}}\right)^4 & \mathrm{a}_{i, \mathrm{osc}}\leq \mathrm{a} \leq \mathrm{a}_m\, , \\[1em]
		
	 \left(\frac{\mathrm{a}_\mathrm{end}}{\mathrm{a}_m}\right)^4\left(\frac{\mathrm{a}_m}{\mathrm{a}_{\,}}\right)^3 & \mathrm{a}_m \leq \mathrm{a} \leq \mathrm{a}_\mathrm{RH}\, , \\[1em]
		 
	 \left(\frac{\mathrm{a}_\mathrm{end}}{\mathrm{a}_m}\right)^4\left(\frac{\mathrm{a}_m}{\mathrm{a}_\mathrm{RH}}\right)^3\left(\frac{\mathrm{a}_\mathrm{RH}}{\mathrm{a}_{\,}}\right)^4  & \mathrm{a}_\mathrm{RH} \leq \mathrm{a} \leq \mathrm{a}_\mathrm{EQ}\, , \\[1em]
		  
	\left(\frac{\mathrm{a}_\mathrm{end}}{\mathrm{a}_m}\right)^4\left(\frac{\mathrm{a}_m}{\mathrm{a}_\mathrm{RH}}\right)^3\left(\frac{\mathrm{a}_\mathrm{RH}}{\mathrm{a}_\mathrm{EQ}}\right)^4\left(\frac{\mathrm{a}_\mathrm{EQ}}{\mathrm{a}_{\,}}\right)^3    & \mathrm{a}_\mathrm{EQ} \leq \mathrm{a}\, ,
		 
	\end{cases}
\end{equation}
where $\mathrm{a}_{\mathrm{EQ}}$ denotes the scale factor at matter-radiation equality,
defined by the condition $\rho_r(\mathrm{a}_{\mathrm{EQ}}) = \rho_m(\mathrm{a}_{\mathrm{EQ}})$,
where $\rho_m$ is the total matter energy density
normalized to reproduce the present-day matter abundance
measured by Planck, $\Omega_{m,0}=0.3111\pm0.0056$~\cite{Planck:2018vyg}.
Since the \ac{DM}
ensemble remains subdominant during reheating, the initial conditions
derived in the previous section provide a consistent starting point for
the late-time evolution.
The individual abundances $\Omega_i$ of each field $a_i$ are thus given at the various epochs
(indicated by the value of the scale factor $\mathrm a$) by
\begin{equation}
\footnotesize
 \frac{\Omega_i}{\chi_i} = \begin{cases}
 \mathrm{a}\exp\!\left[-3\beta_i\, \mathrm{a}_{i,\mathrm{osc}}^2 \left(\frac{\mathrm{a}^2}{\mathrm{a}_{i,\mathrm{osc}}^2}-1\right)
\right]\, ,  & \mathrm{a_{i, osc}}\leq \mathrm{a} \leq \mathrm{a}_m\, , \\[1em]

    \mathrm{a}_m\exp\!\left[-\beta_i \mathrm{a}_{m}^2\left(4\left(\frac{\mathrm{a}}{\mathrm{a}_m}\right)^{3/2}-1-3\left(\frac{\mathrm{a}_{i,\mathrm{osc}}}{\mathrm{a}_m}\right)^2\right)\right]\, , & \mathrm{a}_m \leq \mathrm{a} \leq \mathrm{a_{RH}}\, , \\[1em]

    \left(\frac{\mathrm{a}_m}{\mathrm{a_{RH}}}\right)\mathrm{a}\exp\!\left[-\beta_i\, \mathrm{a}_{\mathrm{RH}}^2\left(\sqrt{\frac{\mathrm{a}_m}{\mathrm{a}_{\rm RH}}}
\left(3\left(\frac{\mathrm{a}}{\mathrm{a}_{\rm RH}}\right)^2+1\right)
-\left(\frac{\mathrm{a}_m}{\mathrm{a}_{\rm RH}}\right)^2
-3\left(\frac{\mathrm{a}_{i,\rm osc}}
{\mathrm{a}_{\rm RH}}\right)^2\right) \right]\, , & \mathrm{a}_\mathrm{RH} \leq \mathrm{a} \leq \mathrm{a_{EQ}}\, , \\[1em]

 \left(\frac{\mathrm{a}_m}{\mathrm{a_{RH}}}\right)\mathrm{a_{EQ}}\exp\!\left[-\beta_i \,  \mathrm{a}_{\mathrm{EQ}}^2\left(
 \sqrt{\frac{\mathrm{a}_m}{\mathrm{a}_{\rm RH}}}\left(4\left(\frac{\mathrm{a}}{\mathrm{a}_{\rm EQ}}\right)^{3/2}+\frac{\mathrm{a}_{\rm RH}^2}{\mathrm{a}_{\rm EQ}^2}-1\right)-\left(\frac{\mathrm{a}_m}{\mathrm{a}_{\rm EQ}}\right)^2
-3\left(\frac{\mathrm{a}_{i,\rm osc}}{\mathrm{a}_{\rm EQ}}\right)^2
\right)
\right]\, , & \mathrm{a}_\mathrm{EQ} \leq \mathrm{a}\, ,
  \end{cases} 
\end{equation}
while for the fields oscillating after $\mathrm{a}_m$,
\begin{equation}
\footnotesize
     \frac{\Omega_i}{\chi_i} = \mathrm{a}_m\cdot \begin{cases}
\exp\!\left[-4\beta_i\,\sqrt{\mathrm{a}_m\mathrm{a}_{i, \mathrm{osc}}^{3}}\left(\left(\frac{\mathrm{a}}{\mathrm{a}_{i,\mathrm{osc}}}\right)^{3/2}-1\right)\right]\, , &\mathrm{a}_{i, \mathrm{osc}}\leq \mathrm{a} \leq \mathrm{a}_\mathrm{RH} \, ,\\[1em]

\left(\frac{\mathrm{a}}{\mathrm{a_{RH}}}\right)\exp\!\left[-\beta_i 
\sqrt{\mathrm{a}_m\mathrm{a}^3_{\mathrm{RH}}}
\left(
3\left(\frac{\mathrm{a}}{\mathrm{a}_{\mathrm {RH}}}\right)^2-4\left(\frac{\mathrm{a}_{i,\mathrm{osc}}}
{\mathrm{a}_{\mathrm{RH}}}\right)^{3/2}
+1
\right)
\right]\, , &\mathrm{a_{RH}}\leq \mathrm{a} \leq \mathrm{a}_\mathrm{EQ} \, ,\\[1em]

\left(\frac{\mathrm{a_{EQ}}}{\mathrm{a_{RH}}}\right)\exp\!\left[-\beta_i\,\mathrm{a}^2_{\mathrm{EQ}}
\sqrt{\frac{\mathrm{a}_m  }{\mathrm{a}_{\mathrm RH}}}
\left(
4\left(\frac{\mathrm{a}}{\mathrm{a}_{\mathrm{EQ}}}\right)^{3/2}
+\left(\frac{\mathrm{a}_{\mathrm{RH}}}{\mathrm{a}_{\mathrm{EQ}}}\right)^2
-4\left(\frac{\mathrm{a}_{i,\mathrm{osc}}}{\mathrm{a}_{\mathrm{EQ}}}\right)^{3/2}\sqrt{\frac{\mathrm{a}_{\mathrm{RH}}}{\mathrm{a}_{\mathrm{EQ}}}}
-1
\right)
\right]\, , &\mathrm{a_{EQ}}\leq \mathrm{a} \, ,
\end{cases} 
\end{equation}
with
\begin{equation}
  \chi_i ~:=~ \frac{ \rho_{i,\mathrm{osc}}\, \mathrm{a}_{i, \mathrm{osc}}^3  }{\rho_\mathrm{end} \,\mathrm{a}^4_\mathrm{end}}\,,
\end{equation}
where the values of the scale factor corresponding to the different
cosmological epochs are listed in \Cref{tab:scalefactor}.
In order to reproduce the observed \ac{DM} abundance today, it is therefore
crucial to determine the appropriate initial energy densities  $\rho_{i,\mathrm{osc}}$  for each
component of the ensemble. Since the late-time abundances are directly
set by the initial field displacements through the misalignment
mechanism, different choices of initial conditions lead to different
contributions to the total relic abundance. Consequently, obtaining the
correct present-day \ac{DM} abundance imposes non-trivial
constraints on the initial conditions of the fields.

\begin{table}[t]
	\centering
	\begin{tabular}{ccccc}
		\toprule
		& $\mathrm{a}_\mathrm{end}$ &
		$\mathrm{a}_\mathrm{m}$ &
		$\mathrm{a}_\mathrm{RH}$ &
		$\mathrm{a}_\mathrm{EQ}$ \\
		\midrule
		$\mathrm{a}/\mathrm{a}_\mathrm{end}$
		& $1$
		& $35.86$
		& $3.97\times10^{21}$
		& $2.99\times10^{28}$ \\
		\bottomrule
	\end{tabular}
	\caption{\label{tab:scalefactor}
        Characteristic values of the scale factor during the cosmological evolution, normalized to its value at the end of inflation, $\mathrm{a}_{\rm end}$. The quantities $\mathrm{a}_m$, $\mathrm{a}_\mathrm{RH}$, and $\mathrm{a}_\mathrm{EQ}$ denote the transition to the quadratic inflaton regime, the end of reheating, and the matter-radiation equality epoch, respectively.}
\end{table}

In practice, this corresponds to selecting suitable initial vacuum
expectation values, or equivalently the misalignment angle $\theta$. 
These initial conditions determine the amount of energy stored in
each component prior to the onset of oscillations and therefore govern
the subsequent cosmological evolution of the ensemble. The observed \ac{DM}
abundance can thus be interpreted as arising from a particular
distribution of initial field displacements across the tower.

According to the latest Planck results, the present-day
\ac{DM} abundance is measured to be
\begin{equation}
 \Omega_{\mathrm{DM},0}h^2 =0.11933 \pm 0.00091\;\;({\small 68\%\text{, TT,TE,EE+lowE+lensing+ BAO,~\cite[Table 2]{Planck:2018vyg}}})\, ,   
\end{equation}
which corresponds to
\begin{equation}
    \Omega_{\mathrm{DM},0}\simeq 0.26
\end{equation}
for the measured value of the Hubble parameter ($h=0.6766 \pm 0.0042$). Our choice of initial
conditions is fixed such that the total relic abundance of the ensemble
matches this observed value at late times. For this we require that
$\theta=\num{6.99e-5}$ in \cref{eq:vevphi0}. This election of course can
change the initial conditions that we chose for inflation, for
these we change the original potential by a global phase in
\begin{align}
\label{eq:V299}
    V(a^{299})&\simeq \sum_{m=0}^{M} \frac{1}{2}m^2 M_c^2 \left(R_{299 m}a^{299}\right)^2 \nonumber\\
    &+ \Lambda^4\left[ 
	\cos\left( \frac{\sum^{M}_{m=0}R_{299m}r_m a^{299}}{f_a} +\Theta +\Psi\right) -\frac{\kappa}{n^2}
	\cos\left( \frac{n\sum^{M}_{m=0}R_{299m}r_m a^{299}}{f_a} + n\,\Psi\right)\right] \, . 
\end{align}
The resulting evolution of the individual abundances is shown in
\Cref{fig:abundances}. As expected in a \ac{DDM} framework, the total
\ac{DM} abundance is distributed among multiple components with
different masses and lifetimes. The figure also illustrates the
transition between the radiation- and matter-dominated eras at
$\mathrm{a}_{\mathrm{EQ}}$, as well as the present-day scale factor
$\mathrm{a}_0=1$. At late times, the abundances asymptote to constant values, and
their sum reproduces the observed \ac{DM} relic abundance measured today.

\subsection{Tower fractions}

\begin{figure}
    \centering
    \includegraphics[width=0.5\linewidth]{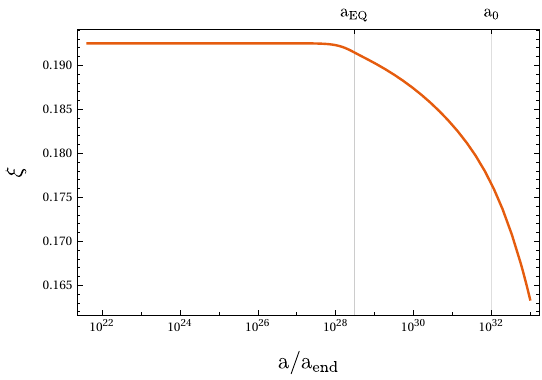}
    \caption{Evolution of the tower fraction $\xi$ as a function of the scale factor
$\mathrm{a}$. The figure shows that $\xi \simeq \mathcal{O}(0.1)$ throughout
most of the cosmological evolution, indicating a significant departure
from the standard single-field \ac{DM} scenario.}
    \label{fig:tfplot}
\end{figure}

Two useful quantities for characterizing the properties of the ensemble
are the tower fraction $\xi$ and the effective equation of
state~\cite{Dienes:2011ja}. The tower fraction quantifies how the total
\ac{DM} abundance is distributed among the different components of
the ensemble. More precisely, it measures the fraction of the total \ac{DM}
abundance not accounted for by the single largest contribution,
and is defined as
\begin{equation}
    \xi ~:=~ 1 - \frac{\Omega_{\mathrm{max}}}{\Omega_{\mathrm{tot}}} \, ,
\end{equation}
where $\Omega_{\mathrm{max}}$ denotes the largest individual abundance
among the ensemble components $\{\Omega_i\}$. Small values of $\xi$
correspond to the conventional single-field \ac{DM} picture, in
which one component dominates the total abundance, whereas larger values
indicate a genuinely multi-component \ac{DM} scenario.

The evolution of the tower fraction is shown in \Cref{fig:tfplot}. For our
benchmark model, we obtain a present-day value
$\xi \simeq 0.18$, indicating a non-negligible distribution of the
\ac{DM} abundance across the ensemble and a clear departure from the
standard single-field scenario.

On the other hand, we can also determine the effective equation of
state of the \ac{DDM} ensemble, which characterizes the collective
behavior of all \ac{DM} components. It is defined as
\begin{equation}
    w_\mathrm{eff}
    ~:=~
    -\left(
    \frac{1}{3H}
    \frac{\dd \ln \rho_\mathrm{tot}}{\dd t}
    +1
    \right) \, ,
\end{equation}
where the total energy density of the ensemble is given by $\rho_\mathrm{tot}=\sum_i \rho_i$.
This quantity provides a convenient
way to describe the global evolution of the ensemble as an effective
fluid. In particular, deviations from $w_\mathrm{eff}=0$ would indicate
departures from the standard cold \ac{DM} behavior due to the
presence of decaying or dynamically evolving components within the
ensemble.

\begin{figure} 
\centering 
\includegraphics[width=0.5\linewidth]{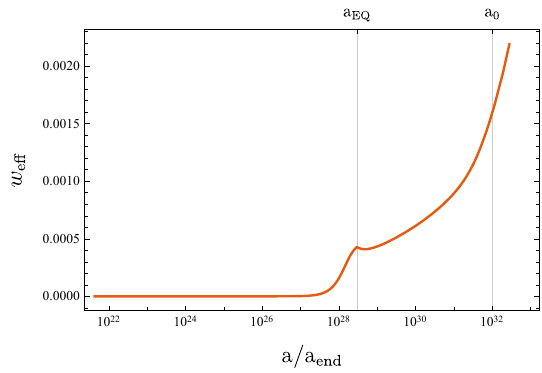} 
\caption{Evolution of the effective equation-of-state parameter $w_{\mathrm{eff}}$
as a function of the scale factor $a$. We find that
$w_{\mathrm{eff}}\simeq 0$ throughout most of the cosmological evolution,
consistent with the expected behavior of \ac{DM}. The peak visible in the
plot arises from the transition in the Hubble expansion rate between the
radiation-dominated and matter-dominated eras, which is modeled here as a
non-smooth transition.} 
\label{fig:wplot} 
\end{figure}

The evolution of $w_\mathrm{eff}$ is shown in \Cref{fig:wplot}. Throughout 
the post-inflationary evolution of the Universe, this effective equation of state 
remains close to zero. This indicates that the field ensemble behaves effectively 
as pressureless matter, despite its multi-component nature, and the 
presence of unstable states. The 
peak at $\mathrm{a}_{\rm EQ}$ is an artifact of our assumption of a non-smooth 
transition for the Hubble parameter $H$ between the radiation-dominated and matter-dominated epochs. At the background level, the present result demonstrates that the 
model successfully reproduces the expected late-time behavior of cold \ac{DM}. At the 
level of fluctuations, a careful evaluation of the resulting matter power spectrum is 
necessary to determine if the non-vanishing value of $w_\mathrm{eff}$ results in a 
cutoff above scales probed by Ly-$\alpha$ forest measurements~\cite{Viel:2013fqw,Narayanan:2000tp,Viel:2005qj,Baur:2015jsy,Irsic:2017ixq,Palanque-Delabrouille:2019iyz,Garzilli:2019qki}. 
We leave this exploration for future work.

An additional aspect of our model concerns the possible generation of isocurvature 
perturbations. In conventional axion \ac{DM} scenarios
produced through the misalignment mechanism, and in particular in the \ac{DDM} 
framework~\cite{Dienes:2012jb}, isocurvature perturbations originate from quantum 
fluctuations of the misalignment angle during inflation. In these scenarios, the \ac{DM} 
sector is typically assumed to be statistically independent of the inflaton field, 
so that the primordial isocurvature fluctuations arise from a degree of freedom distinct 
from the one responsible for inflation. In our case, however, the situation is 
qualitatively different. Only the zero mode $\phi^0$ acquires a non-vanishing \ac{VEV}  
before the onset of non-perturbative effects. This implies that the \acp{VEV} of all 
mass eigenstates are determined by the same misalignment angle $\theta$ through the 
relation given in \cref{eq:vevrotation}. Since this relation also applies to the 
inflaton field $a^{299}$, the initial conditions of both the inflaton and the entire 
\ac{DM} ensemble are determined
by the same degree of freedom, namely the fluctuation $\delta\theta$, implying that 
their primordial fluctuations are not statistically independent but are intrinsically 
correlated. This common origin of the primordial fluctuations distinguishes the present 
framework from conventional axion dark matter scenarios and motivates a dedicated study 
of the resulting isocurvature perturbations, which we leave for future work. 

\section{Final remarks and outlook}
\label{sec:conclusions}

In this work we have studied the cosmological evolution of a tower of axion fields 
arising from the dimensional reduction of a five-dimensional axion model. We have shown 
that, for suitable choices of the model parameters, the framework is capable of 
simultaneously accounting for the inflationary dynamics of the early universe and the 
observed \ac{DM} abundance at late times.

Although the setup contains a large number of axion fields, we found that the 
inflationary dynamics can be described by an effective single-field scenario. 
In particular, the heaviest mass eigenstate, $a^{299}$, dominates the energy density 
during inflation and acts as the inflaton, while the remaining fields remain effectively 
frozen throughout this epoch. 
We further demonstrated that the effective single-field potential yields the same 
inflationary observables as the complete multifield analysis.
This equivalence allows for a much simpler treatment of the inflationary dynamics and 
its associated cosmological perturbations.
Moreover, we showed that our model can also
accommodate the recent ACT-preferred value for the scalar spectral index through a small
shift of the relative phase $\Theta$, while keeping the tensor-to-scalar ratio extremely
small and consistent with current observational bounds.

Following the end of inflation, the inflaton undergoes coherent oscillations and reheats 
the Universe through its decay into photons. We derived the evolution of the inflaton 
and radiation energy densities throughout the reheating epoch and showed that the 
radiation bath is sourced predominantly by the inflaton decay. The resulting reheating 
temperature is found to be $T_{\mathrm{RH}}=5.21\,\mathrm{MeV}$, which remains 
compatible with the lower bounds imposed by \ac{BBN}. Higher reheating temperatures can be
achieved by increasing the hierarchy between $m_\Lambda$ and $M_c$.

An additional observational signature of this scenario arises from the predicted 
\ac{CMB} spectral distortions. In particular, the total \ac{SD} signal exhibits 
percent-level deviations relative to the fiducial 
$\Lambda$CDM prediction over the frequency 
range relevant for future PIXIE-like missions. These distortions remain fully consistent 
with current observational constraints while providing a potential target for the next
generation experiments.

We then investigated the cosmological evolution of the remaining axion fields. Since the 
Hubble parameter during reheating becomes much smaller than the masses of all lighter 
eigenstates, these fields begin oscillating before the onset of the radiation-dominated 
era. Their subsequent evolution is governed by the misalignment mechanism, leading to a 
non-thermal production of \ac{DM}. We showed explicitly that the energy densities associated
with these fields remain subdominant throughout reheating and therefore do not affect the reheating dynamics.

A crucial ingredient of the \ac{DM} phenomenology is the choice of the initial field 
displacements, or equivalently the misalignment angle $\theta$. These initial conditions 
determine the energy stored in each component of the tower and therefore fix the 
present-day \ac{DM} abundance. We found that appropriate values of the misalignment 
angle allow our model to reproduce the observed \ac{DM} relic abundance while 
maintaining consistency with the inflationary dynamics and reheating history.

An interesting feature of our scenario is that the present-day \ac{DM} abundance is not 
carried by a single axion field but is distributed among many components of the tower. 
This is reflected in the non-vanishing value of the tower fraction, which indicates a 
significant departure from conventional single-field \ac{DM} models. At the same time, 
the effective equation of state remains very close to that of cold \ac{DM}, ensuring 
compatibility with the standard cosmological evolution at late times.

The presented interplay between inflationary physics, reheating, and
dynamical \ac{DM} offers a rich cosmological structure that deserves further 
investigation. In particular, future work must include a more detailed
 analysis of isocurvature
perturbations and observational signatures associated with the 
multi-component nature of the axion tower.

\vspace{-5mm}
\section*{Acknowledgments}
This work was partly supported by UNAM-PAPIIT IN117226. MG is supported by the DGAPA-PAPIIT grant IA100525 at UNAM, and a Cátedra Marcos Moshinsky.


\providecommand{\bysame}{\leavevmode\hbox to3em{\hrulefill}\thinspace}

\end{document}